# Smectite phase separation is driven by emergent interfacial dynamics

Michael L. Whittaker,[1,2*] Laura N. Lammers,[1,3] Christophe Tournassat[4], Benjamin Gilbert[1,2]

[1] Energy Geosciences Division, Lawrence Berkeley National Laboratory, Berkeley, California, USA 94720.
[2] Department of Earth and Planetary Science, University of California, Berkeley, California, USA 94720.
[3] Department of Environmental Policy, Science and Management, University of California, Berkeley, California, USA 94720.
[4] Université d'Orléans, Institut des Sciences de la Terre d'Orléans, 45071 Orléans, Cedex 2, France

**Abstract**
Smectite clay minerals have an outsize impact on the response of clay-rich media to common stimuli, such as water imbibition and ion exchange, motivating extensive effort to understand microscopic behaviors resulting from these processes such as swelling and exfoliation. Nonetheless, there is no general consensus about the microscopic forces that govern smectite properties, which are model systems for understanding colloidal and interfacial phenomena more generally. We find that the complex free energy surface arising from the interplay of at least four intermolecular forces and their nonlinear couplings that control local particle-particle interactions leads to dynamic, unstable equilibria between distinct phases. Mechanical disequilibrium arising from osmotic gradients between curved or interacting interfaces drive the dynamic exchange of layers and ions between dense and dilute phases via avalanche transitions that are sustained by thermal fluctuations. We suggest that the surprising interfacial dynamics displayed by smectite minerals, arising from the vastly different mobilities of water, ions and mineral, makes them fundamentally distinct from non-clay minerals because their structures are easily perturbed away from simultaneous chemical and mechanical equilibrium.

## 1. Introduction

Clay minerals are nanominerals that, unlike any other class of minerals, are defined by the size and shape of the particles that comprise them[1]. Nanomineral systems therefore have distinct properties from bulk solids[2] due to the pervasiveness of interfaces[3]. Minerals whose maximum particle dimension is below a certain (e.g., <1 μm) size are the primary component of clay[4], and phyllosilicates are layered minerals whose anisotropic chemical bonding (Fig. 1a, b) contributes to their prevalence in this size regime. Much stronger covalent bonding in the plane of the layer relative to the weaker intermolecular forces in the orthogonal, stacking direction (Fig. 1c) can lead to delamination and exfoliation (Fig. 1d) that expose these minerals to chemical and/or mechanical processes that reduce the layer dimensions.

Phyllosilicates are composed of metal oxide polyhedra that arrange into crystalline layers through oxygen sharing at edges and corners (Fig. 1a, b). Isomorphic and heterovalent substitution of a cation with a lower formal charge than the host lattice (e.g., $Al^{3+} \rightarrow Si^{4+}$) within an oxide polyhedron creates an effective negative structural charge (Fig. 1a, b). Substitutions of this type can occur in sheets that are composed of either octahedral (O) or tetrahedral (T) networks, or both (Fig 1c), and the resulting structural charge is balanced by cations that are located between layers. Strong ionic interactions contribute to strong bonding when substitutions occur in all, or most, unit cells, as they do in micas.

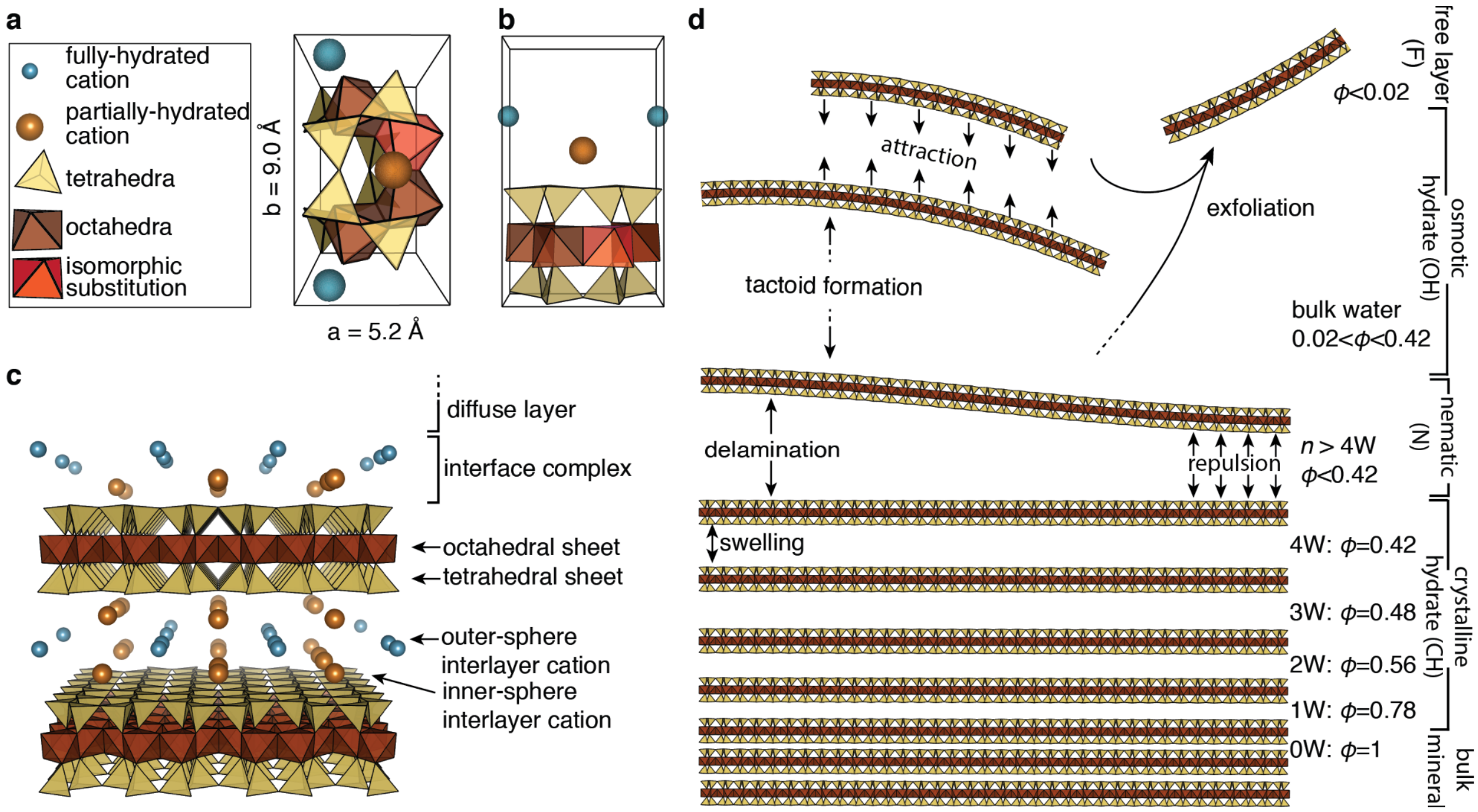

Fig 1 | Hierarchical clay mineral structures of a dioctahedral TOT layer. (a) Fundamental units of clay structures, arranged into unit cell viewed along stacking direction. (b) View of unit cell along plane of layer. (c) Ion configurations at interface with bulk solution and between layers (water molecules not shown for clarity). (d) Common structural motifs observed with varying water activity (varying mineral volume fraction).

Interlayer interactions are weaker when only some unit cells contain substitutions, and water and other molecular species can intercalate between the layers, causing them to swell.

Quantifying attractive and repulsive forces between swelling mineral layers is important because they are intimately connected to important phenomena occurring at both smaller and larger scales. Earth's most abundant interfaces are the surfaces of clay particles[5], which control the behavior of fine-grained rocks because they form a continuous mineral network[6] at low volume fractions, $\varphi$, and change volume substantially as environmental conditions, such as electrolyte concentration, $c$, and water activity, $a_w$, vary[7]. Chemical processes at clay mineral interfaces such as sorption, intercalation, and ion exchange alter the mechanical properties of clay-rich rocks, such as the bulk, storage, and loss moduli. Coupled processes such as these[8,9] mediate environmental transport of carbon and water[10,11], retain $CO_2$ and radionuclides in waste repositories[12,13], concentrate critical elements such as lithium[14] and lanthanides[15] and trigger seismic slip[16–18].

Exfoliated layers or stacks of layers (i.e., tactoids) with high aspect ratios can bend, and naturally occur in a continuum of confirmations that range between the extremes of either planar layers or curved tubes. This conformational continuum is exemplified by TO minerals with the nominal composition $Al_2Si_2O_5(OH)_4$, for which planar kaolinite layers[19] can roll into tubes that are then called halloysite[20] and have higher cation exchange capacities (CEC) and lower bulk moduli[21]. The primary characteristics defining a mineral substance are that it has a unique chemical composition and crystal structure[22], but the ability of an $Al_2Si_2O_5(OH)_4$ mineral to be a non-clay mineral when crystals are sufficiently large, or different types of clay minerals depending on their shape, is one clear demonstration that the size and curvature of clay minerals cause them to behave differently than other mineral substances.

Widely applied definitions of 'clay' include the stipulation that the material is 'plastic at appropriate water contents' and 'will harden when dried or fired'[4]. Phenomenological descriptions such as this leave considerable ambiguity about what exactly can be considered a clay mineral (what if plasticity is observed in the absence of water?), and even *when* (does a clay cease being clay if it transiently softens, instead of hardens, as it dries?). Our interest here is not to parse the semantic aspects of clay mineralogy in order to refine clay taxonomy. Instead, we propose that the variability in definitional nomenclature for clay minerals implicitly identifies *changes* in structure as a key component of the properties of clay, and therefore we suggest that clay minerals, unlike other minerals, must be understood through the time dimension in addition to the spatial dimensions.

We focus here on the swelling clay montmorillonite (Mt), composed of TOT layers that are approximately 1 nm thick and vary in lateral size from just a few unit cells (UC) to microns (Fig. 1a-d). Exchange of molecular species between layers, such as water intercalation during swelling, affects the magnitude and direction of many competing attractive and repulsive interactions over length scales spanning less than an ångstrom to hundreds of nanometers. We describe new experimental results that elucidate the dominant forces controlling microscopic interactions in Mt and present a unified analytical approach for determining how clay systems traverse the free energy surface across environmentally relevant conditions. We first employ LiCl as a representative monovalent electrolyte because its high aqueous solubility allows for the continuous examination of over seven orders of magnitude in electrolyte concentration, approaching the limits of aqueous water activity at high concentrations. We then employ $CaCl_2$ as a representative divalent electrolyte to demonstrate that the same fundamental forces underlie the energetics in this system, but with starkly different consequences.

## 2. Observing colloidal phase separation

Aqueous dispersions of homoionic Mt were prepared over a range of $\varphi$ and $c$. Representative microstructures for $\varphi$ = 2% Li-Mt suspension in water, with no added electrolyte, were characterized with cryo electron tomography (cryoET, Fig. 2a-e). CryoET preserves the fully hydrated mineral *in aqua* and enables visualizing suspension microstructures in three-dimensions over many spatial scales[9]. The importance of tomographic imaging is clearly demonstrated by comparing a two-dimensional cryoEM image (Fig. 2, b) to the 3D reconstructed volume from cryoET (Fig. 2a). Li-Mt has been characterized as a nematic (*N*) liquid crystal, with layers that are aligned along a common axis perpendicular to the plane of the layer, based on projection images such as those in Fig. 2b. However, the curvature of the layers becomes obvious in 3D (Fig. 2c-e). The planar appearance in 2D projection images is a result of optical sectioning of a narrow slice of the curved layer due to strong diffraction contrast from the crystalline lattice when the layer normal is orthogonal to the incident beam direction (i.e., along the zone axis).

Layer curvature has important structural and energetic consequences that are discussed in detail below. Despite being curved, the layers retain an effectively nematic structure, with principal curvature directions aligning along a common axis (Fig. 2d), such that the bulk X-ray scattering patterns (Fig. 2 f,g) are consistent with nematic liquid crystals at $\varphi > 2\%$ and $c < 0.026$ M (Fig. 2h, Fig. S3 and S4). Ensemble structures in both X-ray scattering and cryoET measurements were interrogated via the structure factor, $S(\mathbf{q})$, a measurement of positional order at a scattering vector, $\mathbf{q}$. Radially integrating the three dimensional Fourier transform (FT) of the reconstructed cryoET amplitude yielded an $S(\mathbf{q})$ that enabled deep insights into layered mineral microstructures. Specifically, the cryoET $S(\mathbf{q})$ offers a means to interpret observables such as the average interlayer distance, <*D*>, via direct comparison to the real space 3D reconstruction. In contrast, *in situ* X-ray scattering from bulk suspensions (Fig. 2f-k) cannot be directly inverted from reciprocal space measurements because of the 'phase problem'[23], and therefore cryoET informs the interpretation of X-ray scattering to give both high-resolution and statistical perspectives on clay microstructures.

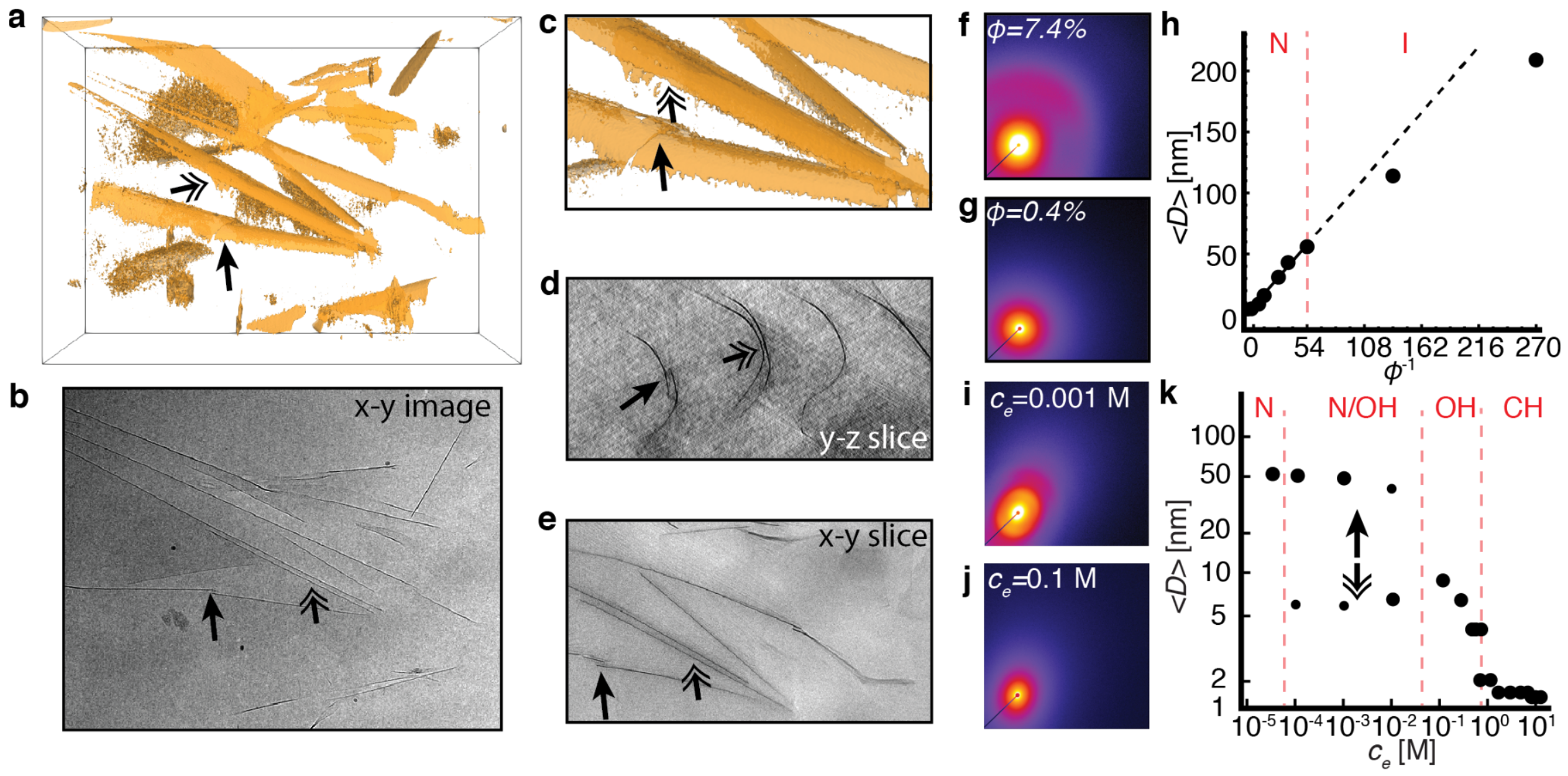

Fig 2 | Smectite microstructures and the effect of electrolyte concentration in LiCl.

$S(\mathbf{q})$ varied strongly with $\varphi$ and $c$ for Mt suspensions in LiCl electrolyte, revealing distinct structural arrangements and, in some cases, coexistence between them. *N* structures are observed when the electrolyte concentration is low or the clay volume fraction is high, but the sharp correlations observed in the *N* structure factor[24] broaden at both lower $\varphi$ or higher $c$, an indication of an increasingly broad distribution of interlayer spacings. This broadening is due to the misalignment of principal curvature axes, which leads to macroscopically isotropic (*I*) arrangements when $\varphi$ < 2% (Fig. 2h). Osmotic hydrates (*OH*), stacks of curved layers with interlayer spacings in the single digit nanometers, coexist with *N* layers at $\varphi$ = 2% and electrolyte concentrations between $10^{-4}$ to $10^{-2}$ M and appear as the primary microstructure between $c = 10^{-2}$ to 0.25 M (Fig. 1g). *OH* structures collapse into crystalline hydrates (*CH*) at electrolyte concentrations above 0.25 M, with the number of interlayer water layers decreasing from three (3W) to one (1W) in discrete steps with increasing electrolyte concentrations.

The transformation of *I* to *N* with increasing $\varphi$, and *N* to *OH* to *CH* with increasing $c$, supports previous observations that coexistence among phases is a common phenomenon in clay mineral systems[8,25]. Our data clearly show that these processes are sensitive to both the clay and electrolyte concentrations which are strongly coupled and, along with the water activity, control the phase behavior of Mt systems. Different forces contribute to the complex energetic landscape in different regions of the composition space, and their consequences for monovalent electrolytes are discussed in section 3. Differences between monovalent and divalent counterions are addressed in section 4, and the energetic and mechanistic origins of smectite dynamics in both cases will be described in section 5.

## 3. Discussion: Quantifying intermolecular forces in monovalent electrolytes

### *3.1 Osmotic potential*

Mt behavior fundamentally arises from structural charge sites within the mineral layers and the counterions that compensate that charge at the mineral interface. The distribution of negative structural charge (Fig. 1a-c) establishes the average layer surface charge density, $\sigma$, which is given by

$$\sigma = \frac{2e}{\sqrt{3}d^2} \quad (1)$$

where $e$ is the fundamental charge arising from a substitution (e.g., $Mg^{2+} \rightarrow Al^{3+}$) and $d$ is the distance between charges in the (pseudo)hexagonal lattices of clay mineral layers (Fig. 1). Lattice parameters of Mt[26] are $a$ = 5.20 Å and $b$ = 9.00 Å and the structural charge is 1.1 $e$/nm[27], giving $d \approx 1$ nm.

Substitutions in Mt are primarily located in the octahedral sheet of the ~1 nm thick layers, and while they occur at discrete lattice sites, the resulting effective charge is distributed among the oxygen ions within the coordination polyhedron (Fig. 1a). Furthermore, the local deviations of the electric field arising from discretized charge sites decay in proportion to the exponential of $-4\pi x/\sqrt{3}d$, where $x$ is the distance normal to the plane of the layer. Therefore, the charge experienced at the surface of the layer, approximately 0.5 nm from the midplane, differs by at most 10% compared to the mean-field expression in Equation (1). At 1 nm from the midplane (~0.5 nm from the surface) the field is within 1% of the value expected from Equation (1). The result of these two charge delocalization mechanisms is that the surface charge density is uniformly distributed across the mineral surface from the perspective of all but the species within two water molecule diameters of the interface.

The Coulomb force on an ion near an infinite, flat, charged surface, $F = ze/2\varepsilon\varepsilon_o d^2$, is independent of the distance, $x$, from the surface (i.e., Gauss' Law). However, entropy arising from mutual counterion repulsion enriches the interface with counterions and controls the decay in ion concentration moving away from the interface towards bulk solution. While interfacial ion complexation and binding is mediated by other forces discussed below, the diffuse, outer portion of the electric double layer (EDL) establishes the interface potential, $\psi_o$, according to the Graheme equation for an isolated EDL

$$\psi_o = \frac{kT}{e} arccosh\left[\frac{ze\sigma^2}{4kT\varepsilon\varepsilon_o c} + 1\right] \quad (2)$$

where $k$ is the Boltzmann constant, $T$ is the absolute temperature, $\varepsilon$ is the relative dielectric permittivity of the EDL and $\varepsilon_o$ is the vacuum permittivity. The counterion charge is given by $z$, and in the presence of aqueous anions, both charge magnitudes impact the resulting interface potential (Fig. S1).

The net field in between two similarly charged layers cancels, and the origin of swelling pressure for infinite, planar layers is given by the osmotic potential due to repulsion of the counterions[28]. It is for this reason that clay structures and interactions are best understood in $N$ regime, which generally occurs at low electrolyte concentrations, when layers are maintained in an oriented configuration by the repulsion of neighboring EDLs (Fig. 3). Long-range osmotic repulsion can extend over 50 nm from the interface (Fig 1d), with a free energy, $W_{osmotic}$, given by

$$W_{osmotic} = 128 c_e \kappa^{-1} tanh\left[\frac{ze\psi_o}{4kT}\right]^2 e^{-\kappa D} \qquad (3)$$

where $D$ is the interlayer spacing and $\kappa^{-1}$ is the characteristic (Debye) length over which the interface potential decays due to screening by background electrolyte, given by

$$\kappa^{-1} = \left(\sum_i \frac{c_i e^2 z_i^2}{kT\varepsilon\varepsilon_o}\right)^{-1/2} \qquad (4)$$

where $c_i$ and $z_i$ are the concentration and charge of electrolyte species $i$. The (repulsive) interaction potential experienced by a layer at a distance $D$ from a neighboring layer is plotted in Fig. 3a for a background electrolyte concentration of $10^{-3}$ M. A maximum repulsive potential of 1 $kT$ per unit cell (UC) occurs as $D \rightarrow 0$, and this potential decreases monotonically as the separation between layers increases.

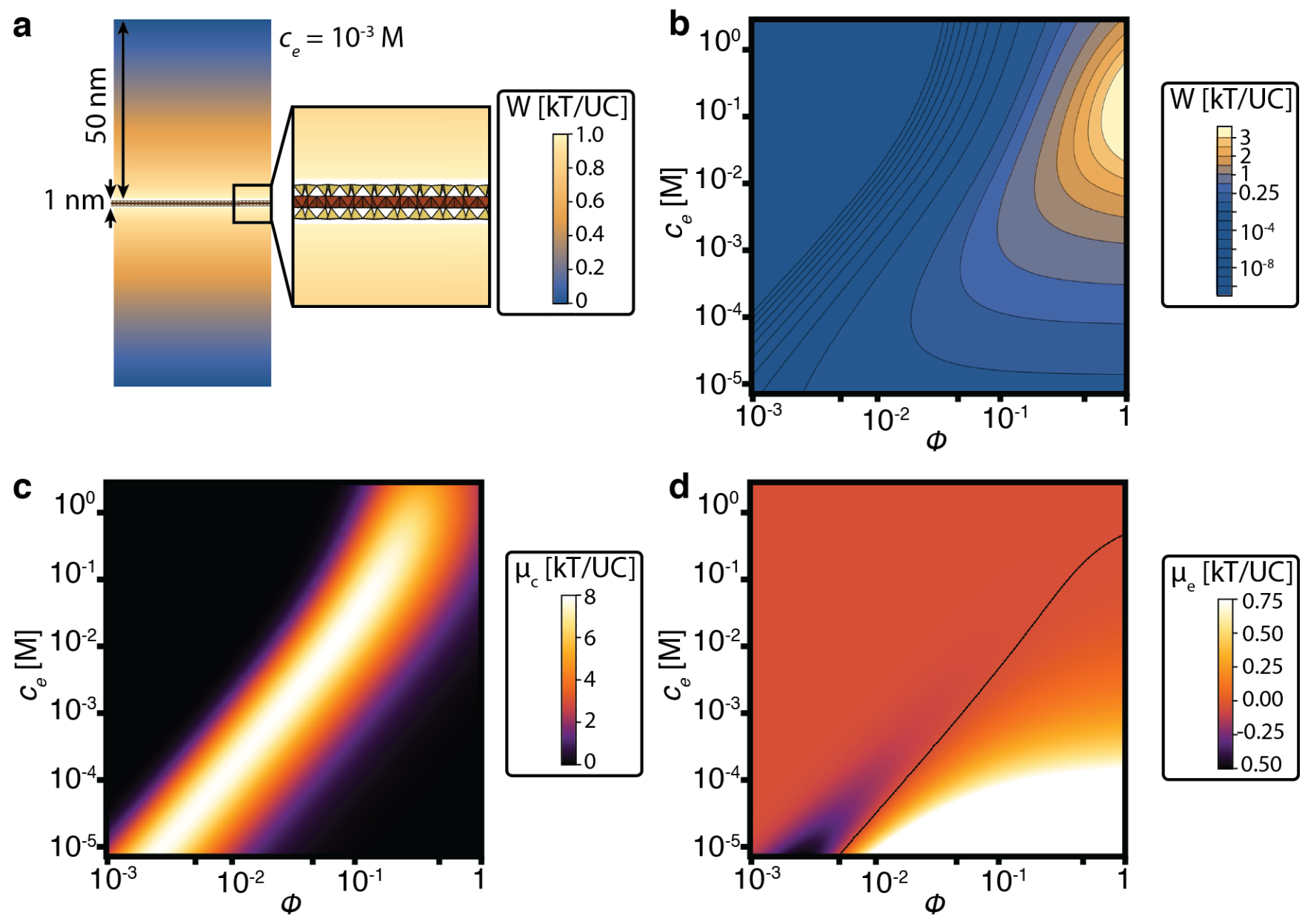

Fig 3 | Osmotic free energy and chemical potentials. (a) Magnitude of $W_{osmotic}$ experienced by a Mt layer at a distance $D$ from a (hypothetical) neighboring layer. (b) Variation of $W_{osmotic}$ with clay and electrolyte concentration. (c, d) Chemical potentials of clay and electrolyte.

From a colloidal perspective, $N$ structures are 'glassy' insofar as their structures are maintained by repulsion[29]. Mineral volume fraction is directly related to the interlayer spacing in the $N$ regime via the relationship

$$D = t/\varphi \qquad (5)$$

Equation (5) holds quantitatively above $\varphi$ = 2% Mt in the absence of background electrolyte (Fig 2h, k). Substituting Equation (5) into Equation (3) gives an expression for the osmotic free energy in terms of the concentrations of both osmolytes, i.e., $\varphi$ and $c$ (Fig. 3b). This interaction energy is globally repulsive (i.e., positive), hence the common description of 'EDL repulsion' in $N$ structures.

The reduction of $\kappa^{-1}$ in the exponential term in Equation (3) with increasing electrolyte concentration is (correctly) implicated in decreasing osmotic repulsion. This is the dominant effect at high clay volume fractions and low electrolyte concentrations[30]. However, osmotic repulsion does not decrease monotonically with increasing electrolyte concentration, and is globally maximized at relatively high electrolyte concentrations (Fig. 3b). This is due to the $\kappa^{-1}e^{\kappa}$ dependence of $W_{osmotic}$ on $\kappa^{-1}$, which arises from the inherent coupling of the mineral and electrolyte osmotic pressures. A maximum osmotic repulsion at intermediate values of $\varphi$ and $c$ (Fig. 3b) means that increasing the electrolyte concentration *increases* the magnitude of the osmotic repulsion at high clay volume fractions, which runs counter to the common assumption that adding salt to a clay suspension monotonically decreases the interaction energy[7]. This non-monotonic change in osmotic pressure with electrolyte concentration has important implications for Mt structures and dynamics by giving rise to phase separated structures.

Chemical potentials of electrolyte and mineral concentrations are given by the gradient of the free energy, $\nabla W$. The chemical potentials of both clay and electrolyte (Fig 3c, d) reveal subtle but important aspects of clay behavior that are hidden by the 'repulsive' nature of $W_{osmotic}$. A local maximum in the clay chemical potential of approximately 8 kT/UC occurs at higher clay volume fractions with increasing electrolyte concentration (Fig 3c). Comparison to the much lower magnitude of the electrolyte chemical potential makes clear that $W_{osmotic}$ is dominated by clay, and not electrolyte, at all but the lowest electrolyte and highest clay concentrations (i.e., the bottom right corner). Only when the electrolyte concentration is very low (<~$10^{-3}$ M) does it have an appreciable impact on the total interaction free energy. Models that only account for the chemical potential of electrolyte and not that of the clay[31][32,33] are applicable only in this regime.

The electrolyte chemical potential is negative when both the clay and electrolyte concentrations are low (i.e., bottom left corner of Fig. 3b-d). In other words, local regions of elevated electrolyte concentration decrease the system free energy. Experimental evidence (Fig. 2h, k) shows that layers undergo an *N* to *I* transition at compositions corresponding to the crossover between positive and negative electrolyte chemical potentials, and this transition is consistent with a transition that causes *N* layers to become misaligned due to the disappearance of globally repulsive forces that maintain glassy *N* structures. An inhomogeneous redistribution of electrolyte around misaligned layers has important impacts on clay mineral structures and how these structures change over time..

Structures are well described by the Percus-Yevick structure factor, $S(\mathbf{q})$, which models layers as ellipsoids that repel via geometric volume exclusion[24], in dilute suspensions. A heuristic relationship for $\langle D \rangle$ based on the harmonic mean of the volume fractions of both flat and oblate ellipsoids has been shown to describe the $S(\mathbf{q})$ across the *N* to *I* transition. The foregoing discussion provides an energetic rationale for this observation: the *N* structure undergoes a transformation to *I* that is driven by the favorable electrolyte chemical potential, but opposed by the unfavorable mineral chemical potential. Thus, bent layers concentrate electrolyte and in the *I* regime but cannot maintain global repulsion, thereby disordering the layer stacking sequences. The *N* structure consists of layers that occupy an effective volume proportional to $\varphi^{-1}$ (Eq. 5), arising from the layer itself and the associated EDL volume stacked along a single axis orthogonal to the plane of the layer. The *I* structure occupies an effective volume[24] proportional to $\varphi^{-1/3}$, consistent with layers that fill three dimensional space, possibly via dynamically bending[9], in response to unstable electrolyte distributions on opposing sides of a layer. *N* and *I* states coexist in proportions that depend on the global value of $\varphi$, which is consistent with the lever rule for phase-separated mixtures[34].

Finally, we note that at low electrolyte concentrations the boundary where $\mu_c = 0$ can be alternatively expressed as the ratio of EDL volume to mineral volume

$$\Phi = \varphi(2\kappa^{-1} + t)/t \quad (6)$$

When $\Phi > 1$, neighboring EDLs overlap and the repulsion strength is given by Equation (3). When $\Phi < 1$, bulk solution is present that is completely screened from the mineral surface charge. Therefore, $\Phi = 1$ marks the limit of $N$ stability because ions within the EDL are out of equilibrium with the bulk solution[35]. Layer disorder is the result of competing attractive and repulsive interactions.

*3.2 Bending*

The bending free energy of a layer, $W_{bend}$, accounts for the elastic energy required to deform the mineral and the excess osmotic free energy of the deformed diffuse ion cloud, both of which are dependent on curvature according to the Helfrich representation[36]

$$W_{bend} = \int \frac{k_1}{2}\left(H_1 + H_2 + H_o\right)^2 + k_2 K \; dS \quad (7)$$

where $k_1$ and $k_2$ are the mean and Gaussian bending moduli, respectively, $H_1$ and $H_2$ are the principal curvatures, $H_o$ is the spontaneous curvature, $K$ is the Gaussian curvature, and the integral is over the layer surface. Principal curvatures can alternatively be expressed as $1/R_i$, where $R_i$ is the radius of curvature of the $i$ axis, and the Gaussian curvature is $1/R_1R_2$. The Helfrich equation is valid in the limit of high electrolyte concentration, when $\kappa^{-1} << R_i$. The first term in Equation (7) is always positive, while the second term changes sign depending on the sign of the Gaussian curvature. The bending moduli are given by

$$k_1 = k_{1,EDL} + k_{1,El} = \frac{\pi\sigma^2\kappa^{-3}}{4\varepsilon\varepsilon_o} + \frac{Et^3}{12(1-\nu^2)} \quad (8a)$$

$$k_2 = k_{2,EDL} + k_{2,El} = -\frac{\pi\sigma^2\kappa^{-3}}{2\varepsilon\varepsilon_o} - \frac{(1-\nu)Et^3}{24(1-\nu^2)} \quad (8b)$$

An infinite combination of principal curvatures can yield the same value of $W_{bend}$ (Fig. 4b) because the layers are not topologically closed and can bend in arbitrary ways in response to changing $\sigma$, $\kappa^{-1}$, $\varepsilon$, and/or $t$. Therefore, the distribution of $R_1/R_2$ is heavily skewed towards curvatures that are vastly different (Fig. 4b). In other words, most layers have one large curvature axis and one axis with negligible curvature, i.e., approximately cylindrical symmetry. This is entirely consistent with observations[9] from cryoET (Fig. 2). Thus, an approximate expression for $W_{bend}$ that neglects the Gaussian curvature, and replaces the integral with the constant surface area of the layer, captures the relevant contribution of the bending energy for isolated layers.

The most important consequence of Equation (7) is that while $W_{bend}$ is always positive for topologically unconstrained layers with minimal Gaussian curvature, its gradient with respect to $c$ and $H$, the electrolyte and curvature chemical potentials, differ on either side of a layer due to the variation in the sign of $H$ (Fig 4 b-d). Thus, the overall energetic effect of curvature is to increase the net repulsion in excess of the EDL repulsion in Equation (3), but the concentration of electrolyte on the concave side of the layer becomes more energetically favorable, while dilution of electrolyte on the convex side becomes equally less favorable.

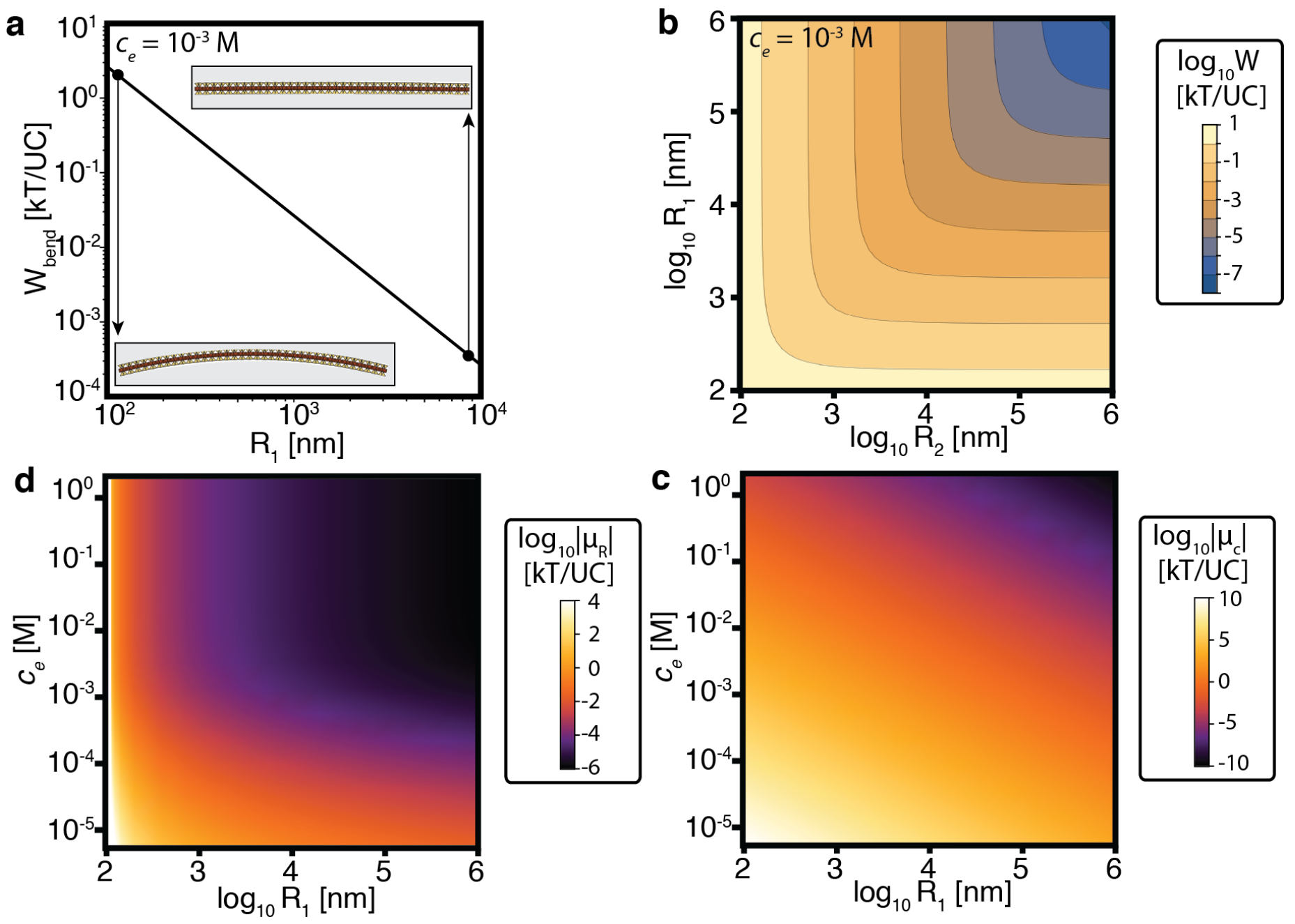

Fig 4 | Bending free energy. (a) Magnitude of $W_{bend}$ for positive and negative curvature. (b) Variation of $W_{bend}$ with the two principal curvature components. (c, d) bending chemical potentials of electrolyte and clay.

Structural observations from cryoET and X-ray scattering data support the existence of an osmotic imbalance across bent layers. The primary consequence is to create an asymmetry in the electrolyte distribution across the layer that arises from the strong dependence of $\kappa^{-1}$ on concentration in Eq. (4). Elevated electrolyte concentrations decrease $\kappa^{-1}$ by a much smaller amount than an equivalent decrease in the electrolyte concentration causes $\kappa^{-1}$ to expand (Fig. S2). Therefore, the total EDL volume of curved layers is greater than that of planar layers at equivalent background electrolyte concentrations. Deviation from Equation (5) towards lower $\langle D \rangle$ observed in Fig. 2h is a result of layers that maintain a net osmotic repulsion but have reduced mean interlayer spacings because their positional order is reduced which may be impacted by dynamic bending (Section 5). Other consequences of curvature, such as the formation of *OH* tactoids, also depend on van der Waals interactions that compete with osmotic repulsion.

### *3.3 Van der Waals*

Derjaguin and Landau, Verwey and Overbeek (DLVO) theory of colloidal interactions describes the competition between osmotic repulsion at large distances, expressed in Equation (3), with van der Waals attraction at smaller distances. Equation (5) can be rewritten to separate the contributions from the thickness of a layer, $t$, and the hydrated interlayer, $D_o$ to the total interlayer spacing, $D$

$$D = \varphi^{-1} t = D_o + t \qquad (8)$$

where $D_o$ is defined as the distance between van der Waals planes at the interfaces of neighboring layers. The van der Waals free energy, $W_{vdW}$, for planar layers is then given by

$$W_{vdW} = -\frac{A}{6\pi}\left[\frac{1}{t^2\left(\varphi^{-1}-1\right)} + \frac{1}{\left(2t\varphi^{-1}\right)^2} - \frac{1}{\left(2t\varphi^{-1}-t\right)^2}\right] \tag{9}$$

where $A$ is the Hamaker constant. The value of $A$ can be determined from Lifshitz theory

$$A = \frac{3kT}{4}\left(\frac{\varepsilon_m-\varepsilon}{\varepsilon_m+\varepsilon}\right)^2 e^{-1/\Phi} + \frac{3h\nu_e}{16\sqrt{2}}\frac{\left(n_m^{\ 2}-n^{\ 2}\right)^2}{\left(n_m^{\ 2}+n^{\ 2}\right)^{3/2}} \tag{10}$$

where $\varepsilon_m$ is the dielectric constant of the mineral, $h$ is Plank's constant, $\nu_e$ is the plasma frequency, $n_m$ is the refractive index of the mineral and $n$ is the refractive index of the solution. All of these quantities are well characterized for clay minerals. The values of $n$ are not expected to vary considerably from their bulk values of 1.55 and 1.3 for clay and electrolyte solution respectively. Previous measurements have shown the plasma energy ($h\nu_e$) to be approximately 25 eV[37], and therefore the Hamaker coefficient is approximately 1.5-2 kT, depending on the electrolyte concentration (Fig 5a).

van der Waals attraction between layers is pronounced at high clay volume fractions, where the magnitude of $W_{vdW}$ grows rapidly over many orders of magnitude approaching $\varphi = 1$ (Fig. 5b). Repulsion interactions originating from hydrated cations and the structured water at the mineral interface, discussed in the next section, also act over similar distances, and the observed behavior of swelling clays is often a result of a competition between these short-range interactions. However, classical DLVO behavior at low clay volume fractions and intermediate electrolyte concentrations resulting from a competition between the osmotic repulsion in Equation (3) and van der Waals attraction in Equation (9) has a distinct signature on clay mineral interactions.

The boundary at which $W_{osm} = W_{vdW}$ aligns quantitatively with the critical coagulation concentration ($CCC$) at which flocculation is commonly reported on phase diagrams. We observe that the

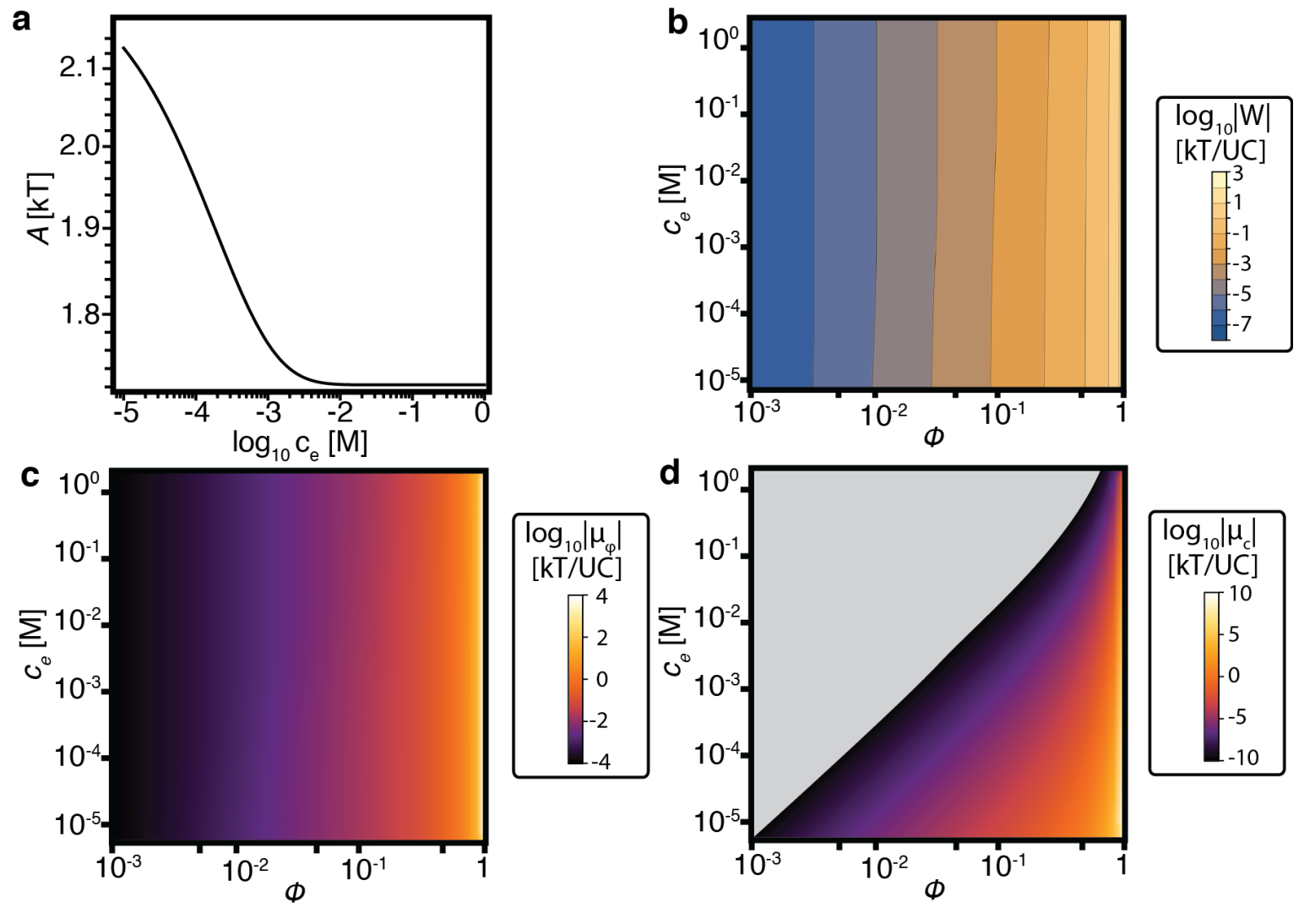

Fig 5 | van der Waals free energy and chemical potential. Variation of the Hamaker coefficient $A$ with electrolyte concentration. (b) Magnitude of $W_{vdW}$, which is minimally impacted by electrolyte concentration. (b). (c, d) van der Waals chemical potentials of electrolyte and clay.

*N* structure separates into coexisting *N* and *OH* (Fig. 2h, k) when the electrolyte chemical potential is negative (Fig. 3d), which is energetically analogous to the formation of the *I* phase, but at higher electrolyte and clay concentrations. Interactions become attractive when $W < 0$, and at $\varphi = 2\%$ only *OH* tactoids are observed starting at 0.026 M (Fig. 2h). This agrees with previous measurements of the *CCC* at the pH values of 8-9 employed in this study[38]. Thus, the formation of *OH* tactoids is entirely consistent with weak but long-range van der Waals attraction[37,39] that is proportionally balanced by $W_{osm}$ that is also very weak.

Osmotic hydrates are stacks of curved layers separated by more than ~2.5 nm of aqueous solution (discussed in section 4.1) in which the layers have an unequal distribution of electrolyte on either side[9]. With reduced charge repulsion and compatible geometry, cylindrical layers can stack into tactoids via attractive interactions, which is distinct from the repulsive interactions that stabilize *N* structures. Decreasing $\langle D\rangle$ with increasing $c$ in the single-phase region of *OH* formation (Fig. 2k) can be quantitatively reproduced with a modified Poisson-Boltzmann expression

$$tan\left[\left(\frac{c_e^{\,n}}{2kT\varepsilon\varepsilon_o}\right)^{1/2}\frac{ze\langle D\rangle}{2}\right]=\frac{\sigma}{c_e^{\,n}}\left(2kT\varepsilon\varepsilon_o\right)^{1/2} \qquad (11)$$

where $n$=1 when layers are planar, while $n = 3/4$ for curved layers. This expression is derived for $n$=1 using the assumption that the repulsive osmotic pressure between the layers is established by the electrolyte concentration at the interlayer midlane, which is in equilibrium with the bulk electrolyte concentration. Sublinear scaling of Equation (11) with the bulk electrolyte concentration (i.e., $n < 1$) is observed above the *CCC* (Fig. 6), and suggests that excess interactions beyond the (ideal) osmotic pressure at the interlayer midplane drive interactions in the *OH* regime.

Excess chemical potential due to curvature arises from counterion distributions that differ from those on planar surfaces, which are solely from the entropy of the counterions (Eq. 3). Specifically, the impact of curvature on interacting layers is that the curvature energy becomes non-additive because the interactions are non-ideal. In the case of two interacting cylinders with radii of curvature $R_1$ and $R_2$ the free energy is given by

$$W_{cyl}=\frac{\kappa^{1/2}}{\sqrt{2\pi}}\left[\frac{R_1R_2}{R_1+R_2}\right]^{1/2}W_{osmotic} \qquad (12)$$

From Equations (4) and (12), $W_{cyl}$ can be seen to scale with $c^{-1/4}W_{osmotic}$, which quantitatively predicts the behavior observed in Fig. 6 at electrolyte concentrations above the *CCC*. Importantly, Equation (13) holds true at both low (Fig. 1g) and high[7] clay volume fractions. This means that a fixed interlayer spacing is adopted regardless of the volume fraction of mineral and its associated EDL when $\Phi < 1$. Thus, the balance between very weak van der Waals attraction given in Eq. 11 and equally weak osmotic repulsion between cylindrical layers in Eq. 12 govern *OH* structures.

Excess osmotic bending energy vanishes when $\kappa^{-1}$ is reduced below the thickness of the interfacial hydration layer (Fig. 4d). For Li-Mt this corresponds to approximately 0.37 nm, or one molecular water layer[33] plus the cation radius, at $c = 0.71$ M (Fig. 1g). This marks the electrolyte concentration at which *CH* tactoids form. A discontinuous jump from $\langle D\rangle = 3.4$ nm to $\langle D\rangle = 1.9$ nm at $c > 0.71$ for Li-Mt cannot be explained by Equations (9) and (13) alone, and hydration forces that act over 1-2 nm from the mineral interface are needed to fully describe clay microstructures in this regime.

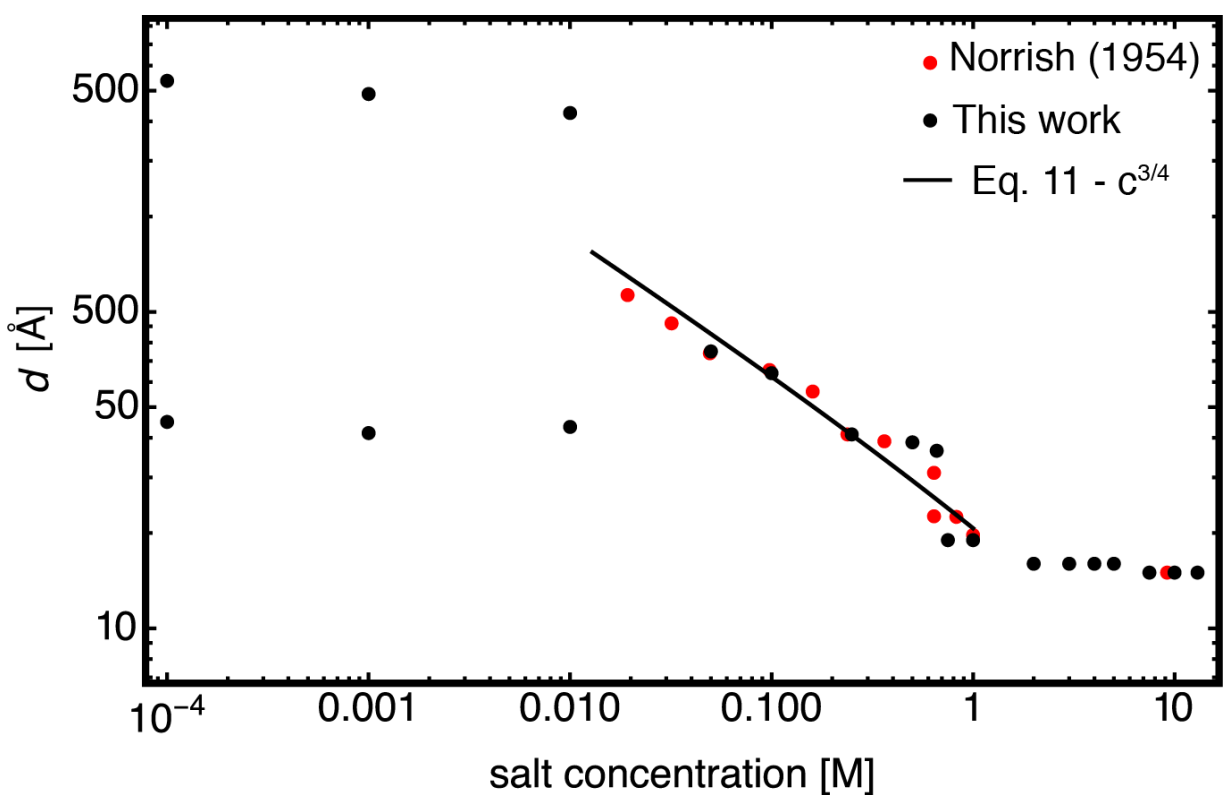

Fig 6 | Interlayer spacing of Li-Mt as a function of added LiCl concentration. Data from Fig. 2k is compared with classical observations by Norrish[7].

*3.4 Hydration complexation*

Water plays a defining role in clay mineral behavior by mediating the binding of counterion-hydration complexes to the mineral surface, which alters the interfacial charge (Eq. 1) and thereby the interfacial potential (Eq. 2). Completely dehydrated Mt (0W) exhibits a basal spacing of approximately 1.0 nm (Fig. 1), the inherent thickness, *t*, of a layer, which includes both the metal oxide framework and the counterions in the interlayer space. Discontinuous transitions to four discrete swelling states each correspond to the addition of a molecular layer of water (~0.3 nm diameter) along the stacking direction and give rise to 1W (1.3 nm), 2W (1.6 nm), 3W (1.9 nm), and 4W (2.2 nm) crystalline hydrates (Fig. 2k, Fig. S3-4).

The total charge concentration on the clay, Σ, can be calculated from

$$\Sigma = \rho t / \sigma Z \qquad (13)$$

where ρ is the dry bulk density and Z is the molar mass. Basal spacings that decrease with increasing charge concentration (i.e., increasing $\varphi$) in the absence of added electrolyte (Fig. S3) predict a basal spacing of 2.0 nm at the maximum value of Σ (i.e., $\varphi=1$, $\rho=\rho_{theoretical}$). This suggests that layers repel more strongly at large distances than what is observed when layers are within a few nanometers of each other. Therefore, the total charge concentration on the mineral must decrease as layers approach, and this is achieved by increased ion binding (i.e., charge regulation).

At $\varphi = 0.05$, the system exhibits *N* order, and there are no (00$\ell$) reflections for $\ell > 1$ (Fig. S4). As the mineral volume fraction approaches $\varphi = 0.1$, a 4W state was observed to co-exist with both 3W and an *OH* with 6.2 basal spacing. Importantly, the interlayer ion configurations differ between the 4W and 3W structures, as evidenced by X-ray scattering peaks associated with (00$\ell$) reflections for $\ell > 1$, indicating that they are spatially segregated and structurally homogeneous. When the mineral volume fraction is increased to $\varphi = 0.2$, the system reverts back to a single *OH* basal spacing of 6.2 nm and a single peak that is consistent with the thickness of a layer, 1.0 nm. Together, this suggests that decreasing the spacing between layers has a strong, and non-monotonic, effect on the interlayer/interfacial cation configuration, even at relatively large distances compared to the radius of the counterion or water molecule.

Ion complexation equilibria are also affected by the electrolyte concentration in solution via mass action. Absence of an X-ray scattering peak corresponding to interfacial ion configurations in *N* structure factors (Fig. S4) indicates that very few ions are bound in fixed inner-sphere positions. This is consistent with the observation that these structures quantitatively obey Eq. 11 (with $n$ = 1, Fig. S5) in which the fully charged interfaces generate a repulsive osmotic potential that establishes the *N* interlayer distance[30,40] described by Eq. 5. Increasing electrolyte concentration promotes the formation of *CH* tactoids, but also reorders interlayer counterion configurations, as observed from shifting (00$\ell$) peaks for $\ell$ > 1. The 1W, 2W and 3W *CH* states form sequentially as the LiCl concentration decreases from 10 M towards zero at $\varphi$ = 0.02 (Figs. 2k, 6), but the initially asymmetric interlayer ion configurations gain the mirror symmetry that gives rise to $\ell$ = even peaks observed in 4W and 3W states (Fig. S4). This indicates that high electrolyte concentrations, which are associated with negative osmotic (Fig. 3) and van der Waals (Fig. 5) electrolyte chemical potentials, promote the concentration of electrolyte at the interface through increased inner-sphere complexation. If ions in the diffuse layer are in equilibrium with those in inner- and outer-sphere positions at the interface, elevating the ion concentration in the diffuse layer is expected to drive increased inner-sphere binding by shifting equilibria, consistent with mass action that is modulated by the crystalline lattice structure of the mineral interface.

Therefore, both $c$ and $\varphi$ affect interfacial ion complexation equilibria, with high concentrations of either component expected to increase the fraction of inner-sphere coordinated counterions and therefore modify the interfacial charge and potential. In the absence of added electrolyte, structured interfacial water on completely exfoliated layers extends approximately 1.5 nm, or the cumulative diameters of four water molecules and two counterions, from the mineral surface[26,41]. Thus, the maximum surface-to-surface separation at which hydration forces are active is approximately 3.0 nm, which corresponds to a basal spacing of 4.0 nm (including the 1.0 nm thick layer itself). This is consistent with the basal spacing of an Na-Mt *OH* tactoid.[7] Small lithium ions are expected to also reside within interstitial gaps between water molecules[42] which is consistent with a 3.4 nm *OH* basal spacing that is predominantly controlled by the molecular diameter of structured water (Fig. 6).

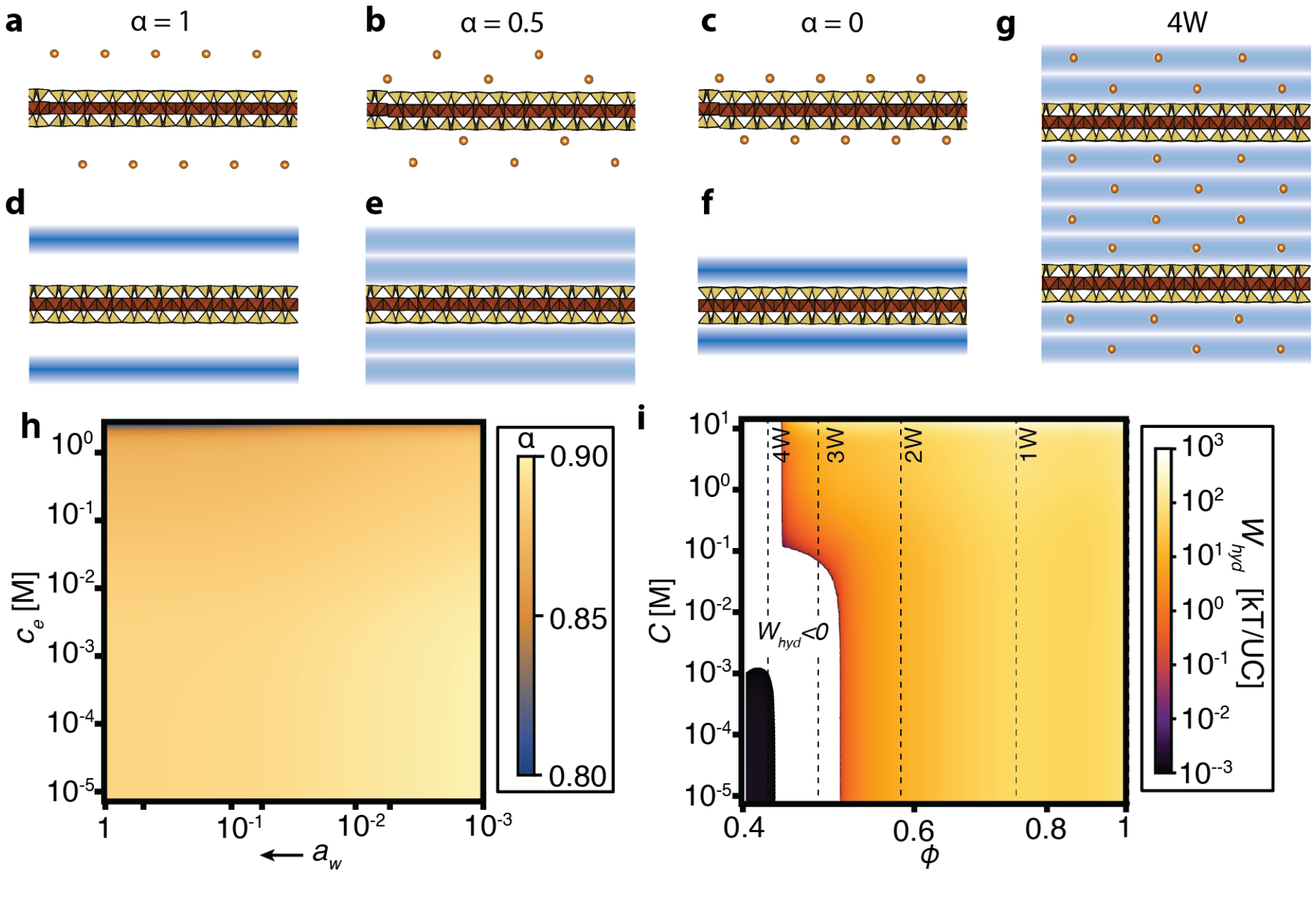

Fig. 7 | Changes in hydration structure with ion complexation. (a-c) Ion configurations at different values of α. (d-f) Hydration profiles corresponding to different values of α. (g) Conceptual representation of a 4W hydrate. (h) Fraction of unbound cations as a function of electrolyte concentration and water activity. (i) Hydration free energy in the high mineral volume fraction regime.

The fraction of ions in outer-sphere complexation configurations, α (Fig. 7 a-c), decreases as electrolyte or mineral concentrations increase, with increased complexation in inner-sphere configurations with increasing $c$ or $\varphi$. The value of α that the system adopts depends on the hydration energy of the cation (Table 1) and the potential drop in the Stern layer, $\psi_s$, which itself depends on the concentration of inner-sphere counterions (i.e., α). Stern potential is a distinct contribution to the total interfacial potential from that given in Eq. 2 (Fig. S6). Importantly, for clay minerals the Stern potential experienced by an ion on one side of a layer is impacted by the Stern potential on the opposite side, and therefore depends strongly on α. When α= 0.5, the Stern potential cancels on each side of a layer, the condition for net electroneutrality under standard state conditions in the absence of ion hydration, for which $a_w = c = 1$.

The cation hydration energy per water molecule[42], $\Delta F_h$, and $\psi_s$ determine the inner-sphere binding energy. Therefore, α can be found for a given water activity, $a_w$, and the electrolyte concentration $c$, via the expression[9]

$$\alpha = \frac{1}{1+\frac{c_e}{a_w}e^{-\Delta F_h + e\psi_s/kT}} \qquad (14)$$

Eq. 14 is a transcendental expression since $\psi_s$ is a strong function of α, but can be solved graphically (Fig S7). Note that $\psi_s$ is strongly dependent on the dielectric constant at the interface, which is poorly constrained. We use measurements of the dielectric constant in high salinity bulk solution (Fig. S8) and the known dielectric constant of the mineral (~7 for Mt) to estimate an interfacial value, which is also a function of α since the dielectric properties of ions and water differ. While we expect that quantifying ion binding energies will require fully atomistic models, qualitative trends predicted by Eq. 14 for reasonably constrained parameters such as the dielectric constant capture the phenomena observed over a wide range of experimental conditions and across different systems[7,24,30,43].

**Table 1. Hydration properties of monovalent cations**[42]

| Ion | Ionic radius (Å) | $CN_w$ | $\Delta F_h$ ($kT/H_2O$) |
|---|---|---|---|
| $Li^+$ | 0.9 | 4.1 | -39.7 |
| $Na^+$ | 1.16 | 5.7 | -31.1 |
| $K^+$ | 1.52 | 6.9 | -25.3 |
| $Rb^+$ | 1.66 | 8.0 | -23.2 |
| $Cs^+$ | 1.8 | 9.6 | -21.5 |

The average hydration potential profile, $W_{hyd}(x)$, is given by Gaussian distributions of water molecules[44,45] centered on the inner- and outer-sphere planes with cation concentrations given by α. At low electrolyte concentrations, an additional plane one water molecule diameter away from the outer-sphere plane supports the formation of 4W because the concentration of counterions in the diffuse layer remains appreciably higher than bulk levels. The total hydration repulsion energy is a convolution of the hydration energy profiles from two adjacent layers. Two overlapping hydration profiles $W_{hyd}(x)$ and $W_{hyd}(-x)$, have a total hydration energy, $W_{hyd}$, given by the inverse Fourier transform of the product of their Fourier transforms, $\mathcal{F}(W_{hyd}) = S_{hyd}$,

$$W_{hyd} = \boldsymbol{\mathcal{F}}^{-1}\left[S_{hyd,1}S^{*}_{hyd,2}\right] \quad (15)$$

where $\boldsymbol{\mathcal{F}}^{-1}$ is the inverse Fourier transform and * denotes the complex conjugate. Using Eq. (2), (3), (4), (14) and (15), the total hydration energy can be calculated for the case of symmetric, planar layers (Fig. 7i).

*3.5 Total interaction free energy in monovalent electrolyte*

Combining the three fundamental interactions between clay layers (osmotic, van der Waals, hydration) with Pauli repulsion between electronic orbitals at very small distances gives the total interaction free energy for clay minerals

$$W = W_{osm} + W_{vdW} + W_{hyd} + W_{Pauli} \quad (16)$$

Note that $W_{cyl}$ in Eq. 12 accounts for the non-additive (non-ideal) effect of curvature on $W_{osm}$ when *OH* layers interact, but is distinct from the contribution of $W_{bend}$ on exfoliated layers or whole tactoids. In general, interactions that alter the potentials between layers based on their geometry can be expressed as a multiplicative correction to the respective term in *W*, acting as a partial activity coefficient. As the degree of curvature increases, it is expected to impact $W_{vdW}$ and $W_{hyd}$ as well[46,47] but this is likely only in the limit of low mineral volume fraction and high electrolyte concentration (i.e., upper left), where other factors are also important.

Coupling between terms, as occurs upon bending, can be included in Equation (16)

$$W = W_{osm} + W_{vdW} + W_{hyd} + W_{Pauli} + W_{bend} \quad (17)$$

Bending adds an additional additive repulsive component that can make the net interaction energy repulsive across most of the free energy surface except within a relatively narrow band below $\varphi = 0.5$ that corresponds primarily to the 4W spacing (Fig. 8b). The free energy surface (*FES*) for smectite swelling (Fig. 8) arising from Eq. 17 is complex, but reproduces experimental observations of structures and behaviors observed across many orders of magnitude in $\varphi$ and $c$. The most important aspect of the *FES* is the sign of the interaction. Positive, repulsive interactions dominate at low $c$ and high $\varphi$, and negative, attractive interactions dominate at high $c$ and low $\varphi$ (Fig. 8a). The reference state for Eq. 17 is pure water. Equilibrium between pure water and the clay-electrolyte system is achieved when the total free energy is equal to zero, i.e., $c$, $\varphi \rightarrow 0$ *or* on the border between net repulsive and attractive interactions (Fig. 8a). The condition $W_{total}>0$ represents suction, for which water uptake is needed to return to equilibrium with pure water. Conversely, when $W_{total}<0$ the addition of pure water concentrates the mineral and drives segregation into dense and dilute states. The formation of the dense state corresponds to aggregation, but it is typically accompanied by the formation of a dilute state as well with which it is in equilibrium. Both are commonly observed in colloidal layered mineral suspensions[40,48–50]. Net attraction can drive gravitational settling, the magnitude of which increases with the increasing mass of aggregated particles. Thus, smectite structures undergo phase separation under a variety of conditions, through which the overall system composition ($\varphi$, $c$) is separated into distinct pairs of ($\varphi_i$, $c_{,i}$) for phases $i$. Gravitational settling is straightforward to visualize on macroscopic scales[40,48–50], and corresponds to left-right movement of $\varphi_i$. More surprisingly, the system can also undergo up-down movement of $c_i$. The mechanism for this is the bending of layers and/or tactoids (Fig. 4), whose energetic variations are primarily due to changes in electrolyte concentration and whose consequence is varying $\psi_s$ and $\psi_o$ on opposing sides of the layer.

The extent to which the system is able to fully segregate depends on the mineral volume fraction, which controls the diffusional mobility (translation and rotation) of the layers. Mineral structures are considered gelled above ~1% due to limited mineral layer mobility[24]. Since the diffusion constants for layers in ~2% solutions are approximately 10 orders of magnitude lower than that of ions and water[9], and this difference is only expected to increase with increasing mineral volume fraction, we assume that the electrolyte generally reaches (local) equilibrium but that the mineral may not. Even for phase separated structures, in which a dilute phase may gain sufficient free volume to fully equilibrate, the dense phase is necessarily concentrated even further above the gel limit. In such cases, a ‘gelled’ structure can be distinguished from ‘glassy’ one if the composition resides in a region with net attraction.

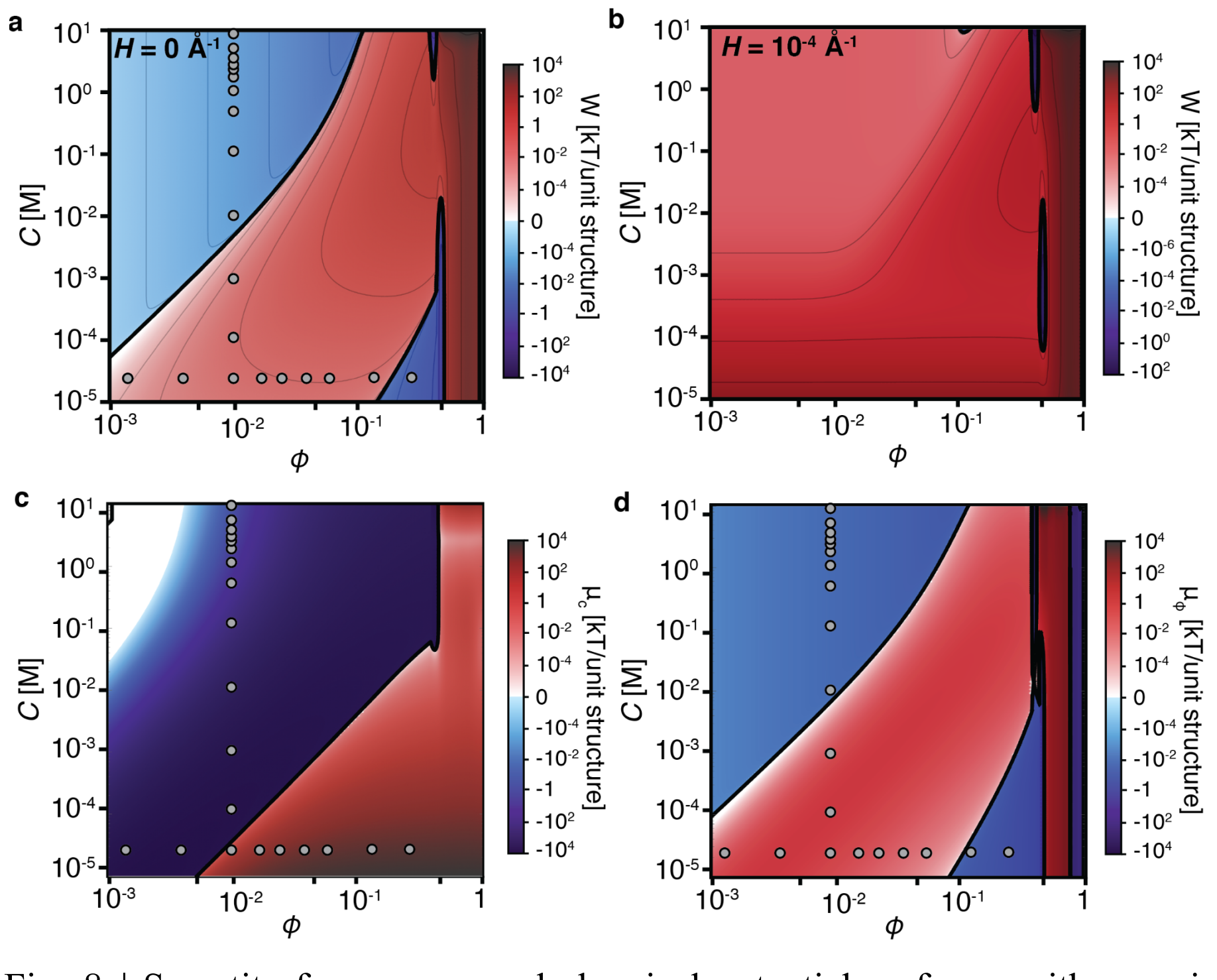

Fig. 8 | Smectite free energy and chemical potential surfaces, with experimental data points shown in gray. (a) Total free energy in the absence of curvature. (b) Total free energy in the presence of curvature. (c) Electrolyte chemical potential. (d) mineral chemical potential.

‘Glassy’ phases can be maintained by mutual repulsion when the chemical potentials of both the mineral and the electrolyte are positive, as they are at low electrolyte concentrations (Fig. 8c,d), This is the origin of the nematic liquid crystalline state observed in Fig. 2 for monovalent electrolytes. Glassy structures are not stable across most of the *FES* because the sign of the chemical potentials differs between electrolyte and mineral, leading to phase separation. Phase separation is a ubiquitous phenomenon arising from the gradient of the *FES,* which is the second most important characteristic governing smectite structures after the sign of the interaction. The gradient of the total free energy, $\nabla W_{total}$, gives the chemical potentials with respect to the electrolyte and mineral concentrations, as well as mineral curvature that couples them

$$\nabla W_{total} = \frac{\partial W}{\partial c} + \frac{\partial W}{\partial \varphi} + \frac{\partial W}{\partial H} = \mu_c + \mu_\varphi + \mu_H \qquad (18)$$

The gradients of the *FES* with respect to $c$ and $\varphi$ are given in Fig. 8c, d. In regions of the *FES* where the chemical potential of a component is positive, local fluctuations that increase the amount of that component increase the free energy. Conversely, when the chemical potential of a component is negative, local fluctuations that concentrate that component are favored. Regions of the *FES* in which the chemical potentials (for each component) are equivalent can coexist. Such coexistence is observed at $\varphi = 0.02$ for $10^{-2} < c < 10^{-4}$, where both nematic and osmotic hydrates are observed, and in other regions of the *FES* at higher $\varphi$ where multiple swelling states (e.g., 4W/3W, Fig. S3) coexist. For smectite suspensions in divalent cations, described below, phase separation is the rule across almost the entire *FES.*

## 4. Quantifying intermolecular forces in divalent electrolytes

### *4.1 Phase separation in divalent electrolyte*

CryoEM images and X-ray scattering of Ca-Mt revealed important structural differences compared to Li-Mt under nominally similar electrolyte and clay concentrations. Ca-Mt tactoids were observed to be phase separated into *CH* and *OH* tactoids (Fig. 9a-e) across all conditions examined. Tactoids with 1.86 nm spacing between layers, corresponding to a 3W hydrate, were separated by an average of approximately 20 nm from distinct tactoids with nearly identical interlayer spacing in the absence of background electrolyte (Fig. 9a). This phase separation, clearly visible in cryoEM images, was also quantified through a strong diffraction ring in wide-angle X-ray scattering associated with the 1.86 nm basal spacing, as well as a broader ring between approximately 20-100 nm, depending on $c$ and $\varphi$ (Fig. 9 f-k, Fig. S4).

At large $c$ and $\varphi$, when the proportion of *CH* tactoids is large, a diffraction ring associated with (002) reflections at approximately 0.98 nm is resolved in wide-angle X-ray scattering (Fig. 9 f-g, i-j). This reflection, as well as higher order reflections up to (008), are observed in cryoEM of a Ca-Mt tactoid for $n$ = 7 layers. The (008) reflection at 0.238 nm is consistent with the approximate diameter of a water molecule, or one molecular layer of water, indicating that the interlayer spacing between Ca-Mt layers is uniform and symmetrical in the limit that the number of layers is large. However, for a tactoid with only three layers this periodicity is lost (Fig. 9d-e). The absence of strong ordering at short distances also corresponds with increased bending that is clear even from the 2D projection image without requiring 3D reconstruction. This suggests that the interlayer configuration becomes increasingly ordered and symmetric as the number of layers increases, which can be at least partly attributed to the decrease in curvature that perturbs interfacial and interlayer structures.

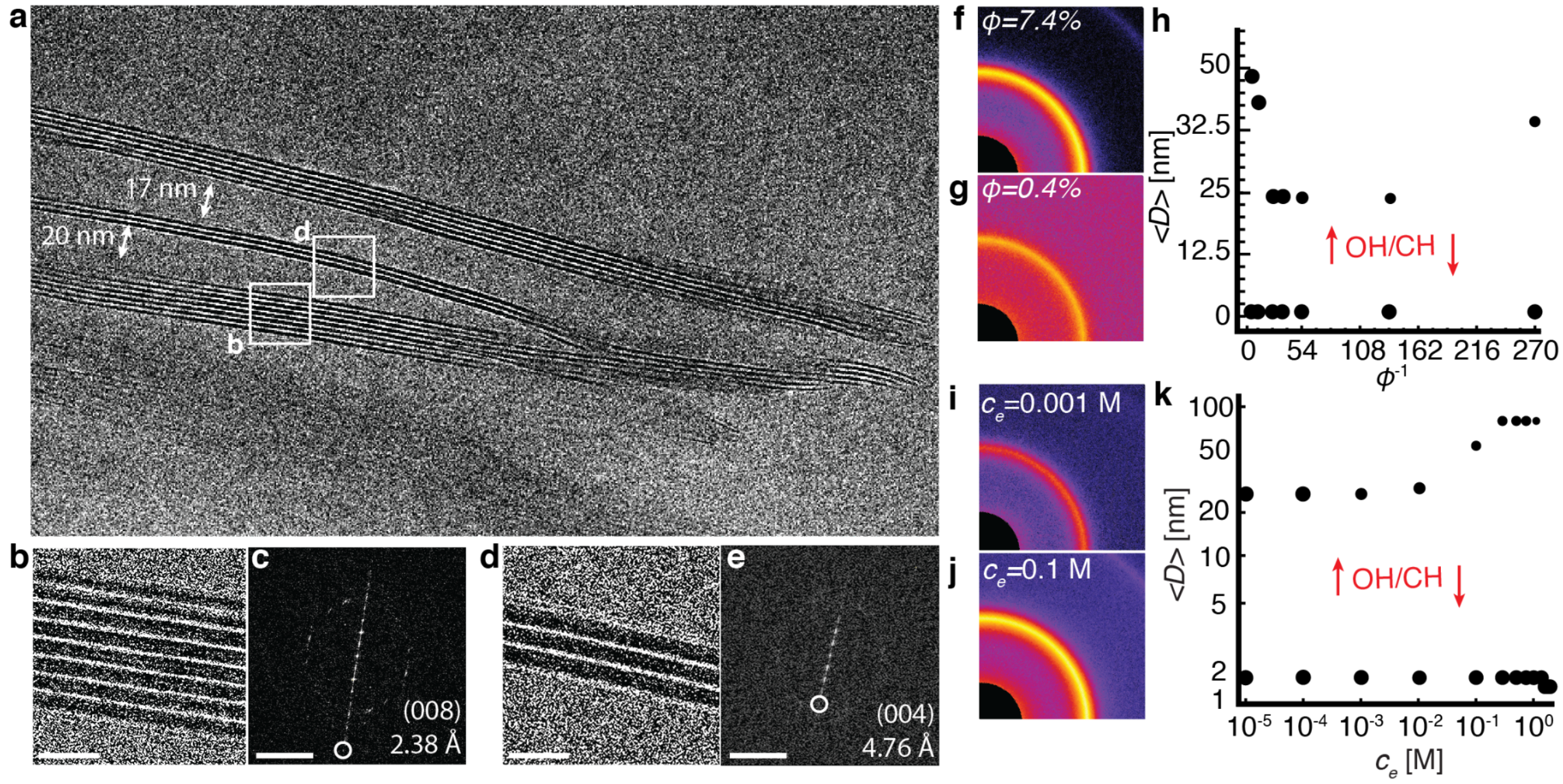

Fig. 9 | Ca-Mt microstructures and the effect of electrolyte concentration in $CaCl_2$. (a) Low-dose cryEM image of Ca-Mt in the absence of background electrolyte. (b) Higher magnification view of seven layer region of tactoid. (c) Fourier transform of (b), showing a minimum (008) spacing along the stacking direction of 2.38 Å, consistent with a plane of water molecules at cryogenic temperatures. (d) Higher magnification view of three-layer tactoid that has higher curvature than the seven layer tactoid in (b). (d) Fourier transform of (d), showing (004) stacking periodicity that begins at 4.76 Å, or approximately half of the thickness of a layer.

*4.2 Interactions in divalent electrolyte*

The four fundamental interactions that control smectite structures in monovalent electrolytes (osmotic, van der Waals, hydration, and Pauli) and their coupling (e.g., bending) are the same as those in divalent electrolytes, but the strong effect of divalent electrolytes on the interfacial potential (Fig. S1) and Debye length (Fig. S2), and their much stronger hydration per $H_2O$ molecule (Table S1) leads to drastically different behavior. Tactoids of *CH* form across the entire range of compositions examined - three orders of magnitude in mineral volume fraction and over five orders of magnitude in electrolyte concentration (Fig. 1h-n). These tactoids also coexist with *OH* tactoids, indicating that phase separation is ubiquitous in Ca-Mt systems. Two important factors control phase separation in divalent electrolytes.

The most important difference between Mt in monovalent and divalent electrolyte solutions is that the interfacial potential of the mineral at a given background electrolyte concentration is much lower (Fig. S1), and for asymmetric (i.e., 2:1) electrolytes, decreases much faster with increasing concentration. The expression for the interfacial potential in 2:2 electrolytes is given by Eq. 2 for $z = 2$. While there is no closed-form analytical expression for 2:1 electrolytes, a solution to the Graheme equation, from which Eq. 2 is derived, can be obtained in this case as well (SI). The interfacial potential inverts (i.e., becomes positive) at 2:1 electrolyte concentrations above approximately 1 M (Fig. S1). By comparison to the case of 2:2 electrolytes, this interfacial potential inversion is due to the presence of the monovalent *anion*, not the divalent cation, which highlights the fact that $\psi_o$ is controlled primarily by the interactions between ions in the diffuse layer and not between the ions and the interface. Additionally, it is important to note that the sign of the potential inverts while the layer remains fully charged, i.e., $Ca^{2+}$ ions are fully hydrated in outer-sphere configurations and not bound to the interface in inner-sphere configurations.

Another important difference between monovalent and divalent electrolytes is that the Debye length (Eq. 4) is much smaller for equivalent concentrations in divalent electrolytes (Fig. S2). Thus, the total volume of mineral and associated EDL (Eq. 6) is much smaller. The combined effect of reduced interfacial potential and reduced Debye length reduces the $W_{osm}$ contribution to $W_{total}$ (Fig. 10a), especially at high electrolyte concentrations. While the topology of the *FES* is qualitatively similar to that in monovalent electrolytes over most of the composition space, the chemical potentials are different than in monovalent electrolytes at high electrolyte concentrations (Fig. 10b,c) leading to distinct phase behavior. Specifically, there is a repulsive regime at elevated electrolyte concentrations for which the magnitudes of the chemical potentials are equivalent to those at much lower concentrations.

Experimental observations show that 3W *CH* tactoids persist across the entire range of $\varphi$ examined in the absence of added electrolyte (Fig. 10). The clay volume fraction in the *CH* is $\varphi = 0.52$, and in order to maintain mass balance, the second, diluted phase must have a mineral volume fraction much lower than the overall starting composition for all values of $\varphi < 0.52$. In other words, the Ca-Mt system must separate along the left-right axis of Fig. 10a-d, and indeed, this is observed.

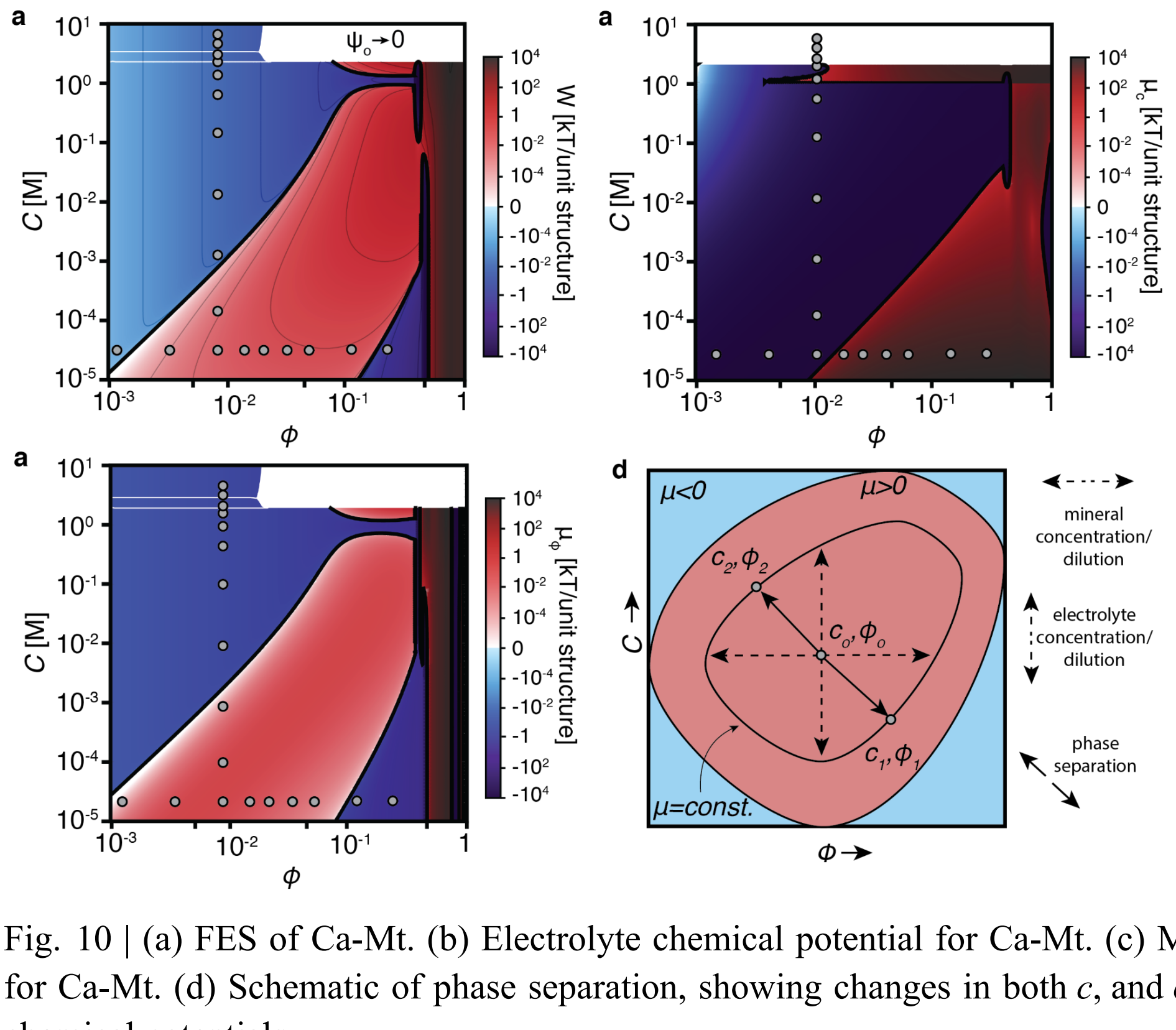

Fig. 10 | (a) FES of Ca-Mt. (b) Electrolyte chemical potential for Ca-Mt. (c) Mineral chemical potential for Ca-Mt. (d) Schematic of phase separation, showing changes in both $c$, and $\varphi$ to states with equivalent chemical potentials.

States for which the chemical potentials of both components, $\mu_{c_e}$ and $\mu_\varphi$, are equivalent can co-exist. However, the chemical potential of electrolyte is positive near $\varphi = 0.52$ and negative below $\varphi = 0.12$ (Fig. 10b), indicating that the electrolyte cannot maintain equilibrium if these phases were to separate along the mineral volume fraction axis alone. The system *can* maintain equilibrium with respect to both components if separation along the up-down (electrolyte concentration) axis occurs simultaneously to separation along the left-right axis, corresponding to changing both $\varphi$ and $c$ in the resulting phases (Fig. 10d). Although electrolyte concentration gradients cannot be maintained in a bulk aqueous solution, concentration of electrolyte in smectite systems can be accommodated when layers can interact and bend. Thus, if the system phase-separates along the top-left to lower-right direction (Fig. 10d), the 3W state may co-exist with dilute *I* or *OH* layers that achieve elevated electrolyte concentrations due to curvature. Importantly, the fraction of these dilute, curved layers can be very low and may go undetected in experiments with insufficient sensitivity or range of length scales.

Despite finding equilibrium when the chemical potentials of each component are equal, if they fall in regions of the *FES* with opposing signs of $W_{total}$ they can create opposing fluctuations that drive competing attraction and repulsion between layers. For example, a Ca-Mt suspension with $\varphi = 0.2$ and $c = 10^{-4}$ M lies in a region with net repulsion ($W > 0$, Fig. 10a) and phase separates into 3W with $\varphi = 0.52$ and $c \approx 10^{-5}$ M, and *I* with $\varphi \approx 10^{-3}$ and $c \approx 1.5$ M. The electrolyte chemical potentials are repulsive (Fig 10b), but the mineral chemical potentials are attractive (Fig. 10c). This leads to inherently dynamic behavior driven by fluctuations in mineral and electrolyte concentrations that favor the formation and disaggregation of tactoids, respectively. Despite the equivalence of the chemical potentials of clay and electrolyte, the *water* is not in equilibrium across phases. Since clay, electrolyte, and water are the only three components in the system, this would appear to violate the Gibbs-Duhem condition that constrains the changes in the chemical potentials. However, curvature acts as an emergent 'component' with a separate chemical potential that maintains distinct water activities in each phase[34]. Curvature is not the only emergent component, as tactoid thickness can also have an analogous effect.

### *4.3 Bending and tactoid thickness*

Similarly to tactoids that form in monovalent electrolytes, Ca-Mt *CH* tactoids collapse from 3W to 2W with increasing electrolyte concentrations (i.e., decreasing water activity). However, 2W was the greatest extent of collapse observed up to the solubility limit of $CaCl_2$ (Fig. 2k). The reason for this is that the relative contribution of $W_{hyd}$ (Table S1) to the total interaction energy is much greater compared to monovalent electrolytes. Thus, hydration is a primary control on the thickness of a tactoid, but the degree to which tactoids curve must also be accounted for.

Tactoid thickness, which enters the expression for $W_{total}$ via the bending modulus in Eq. 8, is primarily affected by curvature. In fact, layer curvature, either due to layer size polydispersity or phase separation, is the reason that tactoids are not infinitely thick. Variation in structural orientation for *N* structures, in which liquid crystalline order is maintained by long-range (glassy) repulsion, is more appropriately quantified using the director field[51] because the tactoid thickness is effectively infinite or changes gradually on the scale of individual layers (Fig. 2). In contrast, tactoid thickness is effectively discontinuous because curved layers with finite sizes are not perfectly geometrically compatible.

This explains why the interlayer separation for the *OH* tactoids that coexist with *CH* tactoids of Ca-Mt is highest at both low and high mineral volume fractions, and lowest at intermediate volume fractions (Fig. 9h, k). The *OH* basal spacings and *CH* thickness are both controlled by the non-monotonic change in osmotic pressure with changing mineral volume fraction (Fig. 10a). This is so because of the tradeoff between osmotic and elastic bending energy, captured in Eq. 8, in which the two terms that contribute to the bending modulus are proportional to $\kappa^{-3}$ and $t^3$, respectively. Including the gradient of the free energy with respect to thickness in the expression for the chemical potentials makes explicit a second emergent system component.

$$\nabla W_{total} = \frac{\partial W}{\partial c} + \frac{\partial W}{\partial \varphi} + \frac{\partial W}{\partial H} + \frac{\partial W}{\partial t} = \mu_c + \mu_\varphi + \mu_H + \mu_t \quad (19)$$

The thickness of the metal-oxide framework of a layer is well constrained (Fig. 9a). Variability in the 'thickness' arises from the size of the interlayer cations, their binding configuration (i.e., α) as well as the water content between layers. The thickness of a tactoid therefore varies discretely as an ion or layer is exchanged. However, because the *FES* varies smoothly with $\varphi$ via Eqs. (5) and (6), with $H$ in Eqs. (7) and (8) and with α via Eq. (14), we expect that thickness can be properly interpreted as a continuous variable that reflects both binding and phase equilibria that are achieved when Eq. (19) is minimized. Furthermore, the 'thickness' of a single tactoid may not be a well defined parameter because stacked and curved systems are likely not static, and different parts of a stack of layers may experience different chemical potentials (e.g., Fig. 9a-e) despite the system residing near an energy minimum of the *FES*.

**5. Dynamics arising from unstable equilibria**

Layer dynamics were characterized using X-ray photon correlation spectroscopy (XPCS) of coherent X-ray scattering in the 42-393 nm size regime, producing speckle patterns corresponding to the dimensions of individual Mt layers[8]. Both autocorrelation of an initial coherent X-ray scattering speckle pattern (one-time) and correlations between speckle patterns collected at times $t_1$ and $t_2 = t_1 + \tau$ (two-time) reflect the relative motion of layers over time.

Two-time correlations for Ca-Mt in 1 M $CaCl_2$ indicate that the layers are mobile on the timescale of seconds, and their mobilities vary with size (Fig. 11 a,b). The near equivalence of one-time and two-time correlations (Fig. 11 c,d) across all sizes investigated indicates that the average fluctuations are not changing appreciably over time, consistent with equilibrium dynamics. Relating the average relaxation time determined from $g_2$ to the square of the scattering vector yields a linear relationship for length scales down to approximately 30 nm, from which the diffusivity of these scattering species can be determined (Fig. 11e). Importantly, this abrupt transition in layer dynamics corresponds directly to the typical spacing of an OH tactoid in these systems (Fig. 9h, k). Thus, the repulsive osmotic force between tactoids suppresses dynamic fluctuations at smaller length scales.

The abrupt nature of transitions between structures, and their return to correlated states after becoming partially uncorrelated, is inconsistent with Brownian diffusion. Stochastic fluctuations lead to correlations that decay exponentially[52], while the Ca-Mt data are best fit with stretched exponentials with a stretching exponent of ~0.3. Furthermore, the relaxation time is essentially independent of the scattering vector at length scales below 30 nm, suggesting that local dynamics are essentially quenched. This suggests that dynamics at small scales are dependent on the diffusive dynamics at larger scales, which is consistent with layers that dynamically bend in opposing directions and for which the electrolyte on either side equilibrates much more rapidly than the layer itself.

Combining insights from structural, energetic, and dynamic data, we conclude that layer curvature produces mechanical imbalances that drive dynamic bending through direct coupling with ion concentrations in solution via the interfacial charge. As shown in Eq. 6, electrolyte in bulk solution that is screened from the interfacial potential of the mineral is not in equilibrium with electrolyte in the EDL. This drives the bending of layers, which increases the electrolyte concentration on the concave side, slightly reducing the Debye length locally (via Eq. 4), and decreases the electrolyte concentration on the convex side, dramatically increasing the Debye length. Thus, the system eliminates 'bulk electrolyte' by increasing the total volume of the EDL such that all ions experience the interfacial potential of the mineral. For large radii of curvature, the energetic penalty for bending is equivalent on either side of a layer (Eq. 7), but the chemical potentials of the components are not equivalent because they depend on the sign of the curvature. Therefore, there is a local osmotic imbalance across layers that is established by the minimization of the total system free energy.

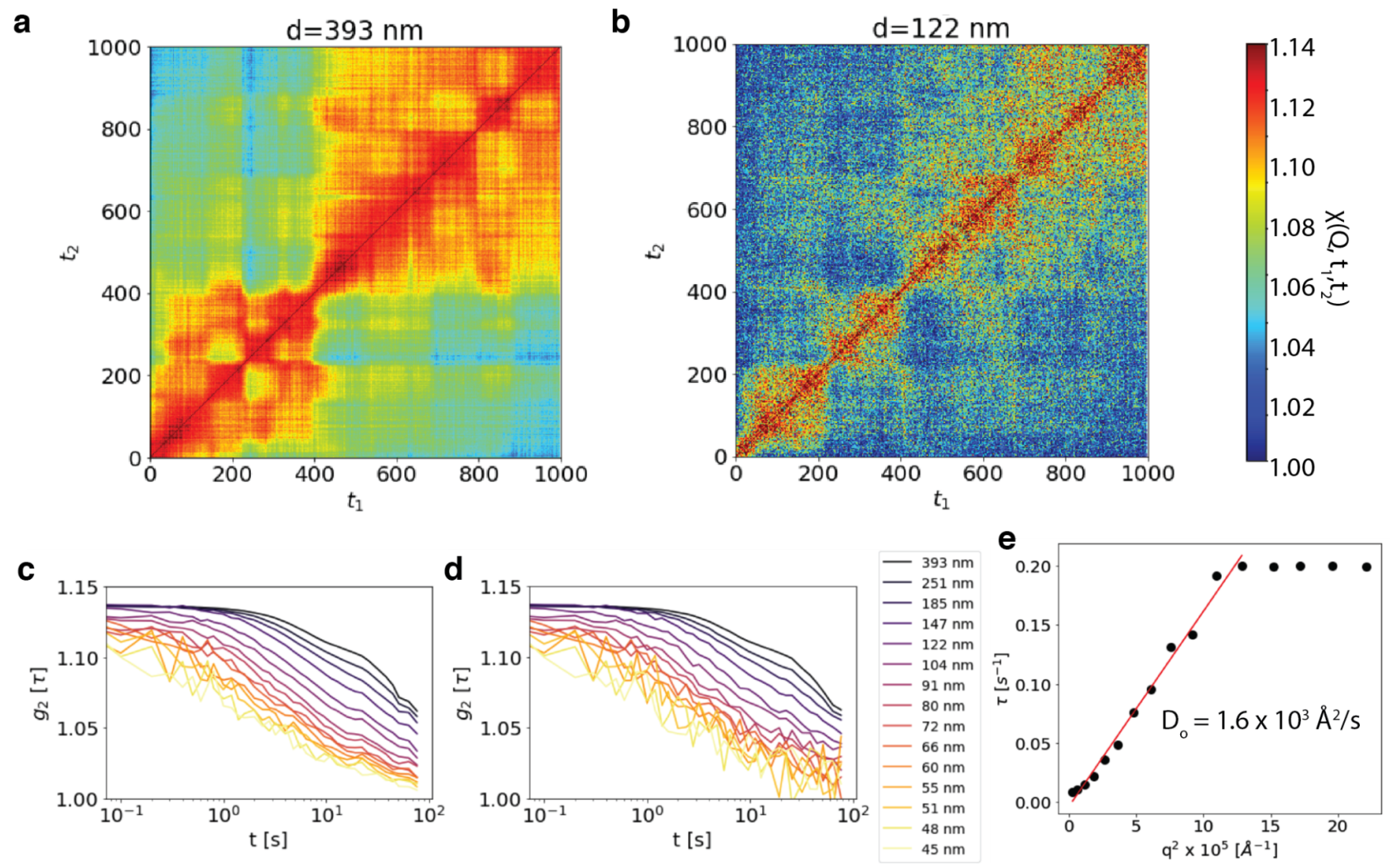

Fig. 11 | Dynamics from XPCS of Ca-Mt systems in 1 M $CaCl_2$ electrolyte. (a) Time-time correlation at $q$ = 0.0016 Å$^{-1}$. (b) Time-time correlation at $q$ = 0.0056 Å$^{-1}$. (c) $g_2$ correlation function from one-time data, corresponding to horizontal axis starting at t=0. (d) $g_2$ correlation function from two-time data, corresponding to profiles taken orthogonal to the primary diagonal. (e) Diffusion coefficient calculated from the dependence of the relaxation time constant, determined from (c) and (d), on the scattering vector.

## 6. Conclusions

Smectite systems consist of three basic components: mineral, water, and electrolyte. Only two of these can be specified independently, and here we chose mineral (represented as a volume fraction, $\varphi$) and electrolyte (represented as a concentration, $c$). Curvature (given by the mean curvature, $H$, or equivalently, radius of curvature, $R$) and tactoid thickness ($t$), are two additional parameters that couple the interaction energies between mineral and electrolyte. Together, variation in the system energy ($W_{total}$) across the free energy surface (*FES*) with changes in $\varphi$, $c$, $H$, and $t$, describe swelling in smectite systems across many environmentally and technologically relevant conditions.

The most important aspect of the *FES* is the sign of the interaction energy, which determines whether interactions are governed by net attraction or repulsion. For a system in equilibrium, the sign of the interaction determines the saturation state relative to pure water vapor at a specified relative humidity (i.e., water activity). The gradient of the *FES* with respect to $\varphi$, $c$, $H$, and $t$ give the chemical potentials for these components. While only the first two are physical components (i.e., mineral or ions), the latter two act as components that emerge as a result of extra degrees of freedom afforded by highly anisotropic layers with much larger length scales than either electrolyte or water. Strictly speaking, $\varphi$ is also an emergent component, the result of collective covalent and ionic interactions that comprise the layer structure (Fig. 1), and the extension of the concept of the chemical potential to curvature or thickness is in principle no different.

The non-additivity of microscopic interactions proposed for some synthetic nanomaterial systems[53] is avoided when accounting for explicit ion binding, its effect on the hydration energy, and the coupling of various parameters arising from the chemical potentials of emergent components. These necessarily lead to collective interactions among clay particles that influence their dynamic equilibration upon phase separation. Nonetheless, we find that accounting for four well-understood interactions (osmotic, van der Waals, hydration, and Pauli) and the coupling between (bending, thickness) them are sufficient to quantitatively recapitulate structures and behaviors in smectite systems, without invoking other attractive interactions such as ion correlation forces[54–56].

The intricate balance between many competing forces can produce non-monotonic changes in properties. These interactions are often inferred from ensemble measurements of bulk properties such as the average basal spacing, <*D*>, and mapped onto phase diagrams[40,43,49]. However, above $\varphi \approx 1\%$ layers are kinetically inhibited from fully sampling the possible space of configurational states due to limited translational and rotational mobility [24]. Thus, $\varphi > 1\%$ for the vast majority of relevant water activities on Earth, and smectites are out of true, stable, equilibrium because layers are forced into contact in unfavorable orientations. These non-equilibrium pairwise layer-layer interactions underlie the time-dependent properties of clays. While chemical potentials in distinct phases may be equal, specifically with respect to mineral and electrolyte, the system may not be in equilibrium with all components (i.e., water). Osmotic gradients manifest as mechanical disequilibrium, causing layers to bend and altering the microscopic chemical forces that are sensitive to curvature. Thus, the coupling between mechanics and chemistry in smectite systems is embodied in the coupling between $\varphi$, $c$, $H$, and $t$ captured in Eq. (19)

Hydrated smectite in dilute suspension exhibits non-ergodic structures, forming glasses or gels that continuously evolve over time[29]. Both glasses and gels of layered minerals exhibit isotropic symmetry and elastic behavior at low mineral fractions, but the defining characteristic of a glass is that the dominant microscopic interaction force is repulsive, whereas a gel is formed in the presence of a net attraction that creates a percolated network[49]. Both types of structures are observed in clays, can interconvert readily, and can be 'rejuvenated' through the addition of mechanical, chemical, or thermal energy[29,50]. We believe that these properties fundamentally distinguish clay minerals from non-clay minerals, the definitions of which do not capture the dynamic behavior that smectites are shown here to exhibit.

## Materials and Methods

**Materials.** Wyoming montmorillonite (SWy-3), obtained from the Source Clays Repository of The Clay Minerals Society (http://www.clays.org/sourceclays_data.html), was used throughout this study. Aqueous solutions of LiCl, NaCl, KCl, RbCl, and CsCl were prepared from reagent-grade salts.

**Cryo-transmission electron microscopy.** Suspensions of Li-MMT or Ca-MMT with particle concentrations of 50 mg/mL were deposited as 2-3 µL aliquots onto 300-mesh lacy carbon Cu grids (Electron Microscopy Sciences) which had been glow-discharged in air plasma for 15 seconds. Excess solution was removed by automatic blotting (1 blot for 10 s, blot force 10 at 100% relative humidity) before plunge-freezing in liquid ethane using an automated vitrification system (FEI Vitrobot). Imaging was performed with a Titan Krios TEM operated at 300 kV, equipped with a BIO Quantum energy filter. Images were recorded on a Gatan K3 direct electron detecting camera in correlated double sampling superresolution mode with a physical pixel size of 1.26 or 0.91 Å/pixel. Imaging was performed under cryogenic conditions using a low electron dose rate ($< 10\ e^{-}Å^{2}s^{-1}$) and low total dose ($< 2000\ e^{-}Å^{2}$) to minimize sample alteration by the electron beam. Acquisition was automated with SerialEM software.

**X-ray scattering.** X-ray scattering was performed at beamline 5ID-D of the Advanced Photon Source at Argonne National Laboratory in order to obtain high photon fluxes necessary for time-resolved experiments. Small-, medium-, and wide-angle X-ray scattering (SAXS/MAXS/WAXS) was collected simultaneously on three Rayonix charge-coupled device (CCD) detectors with sample−detector distances of 8505.0, 1012.1, and 199.5 mm, respectively. The wavelength of radiation was set to 1.2398 Å (10 keV), resulting in a continuous range of scattering vector, $q = 0.017−4.2$ Å$^{-1}$.

**XPCS.** XPCS experiments were conducted at the Coherent Hard X-ray (CHX) beamline 11-ID at the National Synchrotron Light Source II (NSLS-II), Brookhaven National Laboratory. The X-ray energy was 9.65 keV (λ= 1.285 Å) with energy resolution DE/E »$10^{-4}$ from a Si111 double crystal monochromator. A partially coherent X-ray beam with a flux at the sample of ~$10^{11}$ photons/sec and a focused beam size of $10 \times 10$ µm$^2$ was achieved by focusing with a set of Be Compound Refractive Lenses and a set of Si kinoform lenses in front of the sample. The sample was loaded in a wax-sealed glass capillary mounted on the sample stage. The coherent scattering pattern was recorded in transmission small angle scattering geometry by using a photon-counting pixelated area detector (Eiger X 4M Dectris Inc.) located 16.03 meters away from the sample with a 75 µm × 75 µm pixel size. The X-ray radiation dose on the sample was controlled by a millisecond shutter and filters of different thickness of silicon wafers. The data acquisition strategy was optimized to ensure that the measured dynamics and structure are dose independent. The XPCS data analysis were conducted by using software developed at CHX, NSLS-II. A $q$ range of $Q = 0.0015 – 0.9$ Å$^{-1}$, corresponding to length scales of 0.7-420 nm, captures the lateral dimensions and interlayer spacings of all layers[21].

The two-time correlation function, $c(Q, t_1, t_2)$, equals unity if there is no correlation between X-ray scattering intensities at a scattering vector $Q$ for an initial time $t_1$ and second time $t_2$, and approaches 1.2 for intensities that are unchanged between the two time points. The layer diffusion coefficient extracted from a linear fit of $c$ at different $Q$.

**Acknowledgements**

This research was supported by the U.S. Department of Energy, Office of Science, Office of Basic Energy Sciences, Chemical Sciences, Geosciences, and Biosciences Division, through its Geoscience program at LBNL under Contract DE-AC02-05CH11231. Portions of this work were performed at the DuPont-Northwestern-Dow Collaborative Access Team (DND-CAT) located at Sector 5 of the Advanced Photon Source (APS). DND-CAT is supported by Northwestern University, E.I. DuPont de Nemours & Co., and The Dow Chemical Company. This research used resources of the Advanced Photon Source, a U.S. Department of Energy (DOE) Office of Science User Facility operated for the DOE Office of Science by Argonne National Laboratory under Contract No. DE-AC02-06CH11357. Data was collected using an instrument funded by the National Science Foundation under Award Number 0960140. We thank Dan Toso for technical assistance with the FEI Titan Krios. We thank Chenhui Zhu

*To whom correspondence should be addressed mwhittaker@lbl.gov

**Supporting Information**

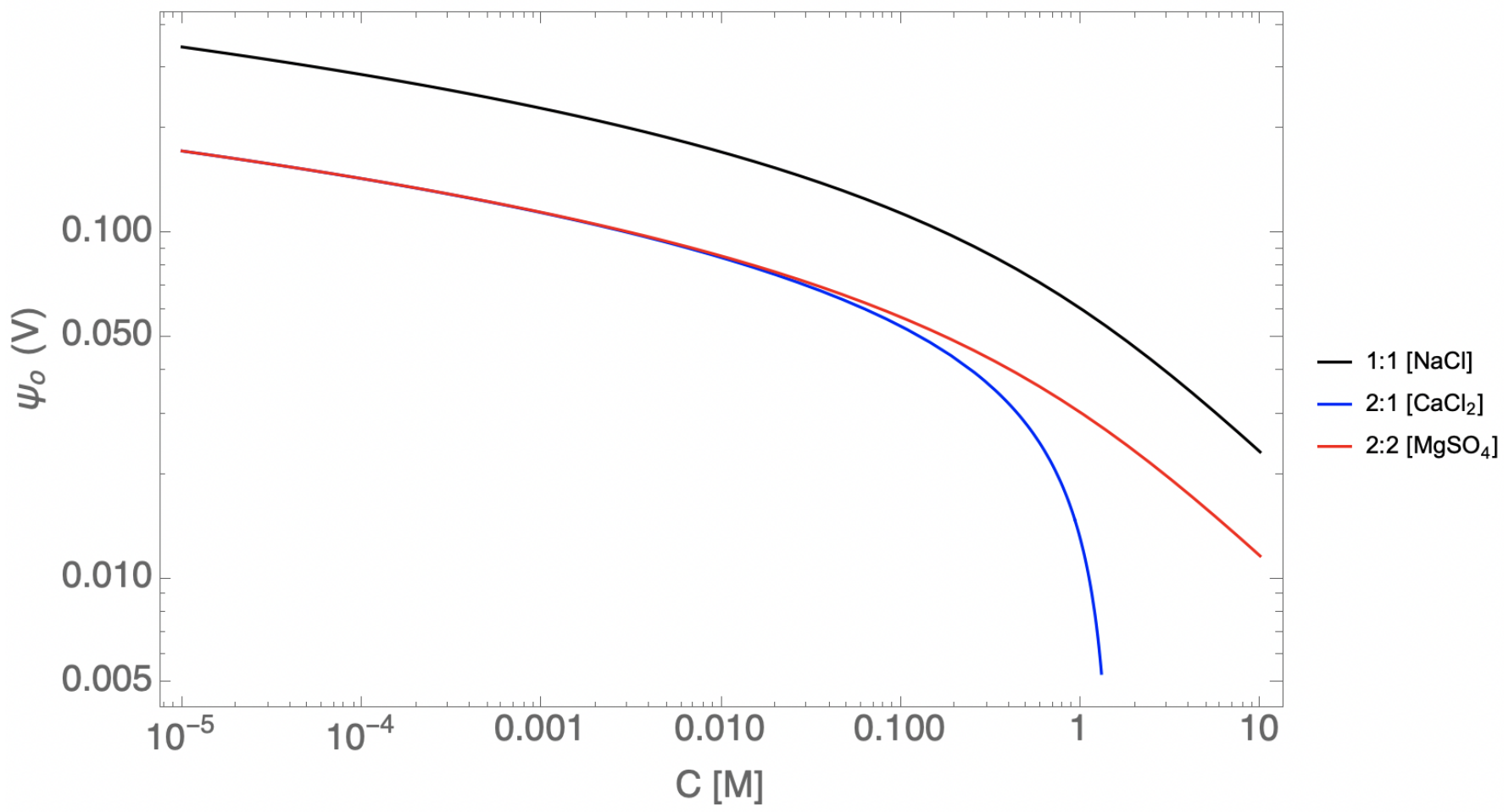

Fig S1 | Interface potential as a function of electrolyte concentration given by the Graheme equation (Eq. 2) for 1:1, 2:1 and 2:2 electrolytes.

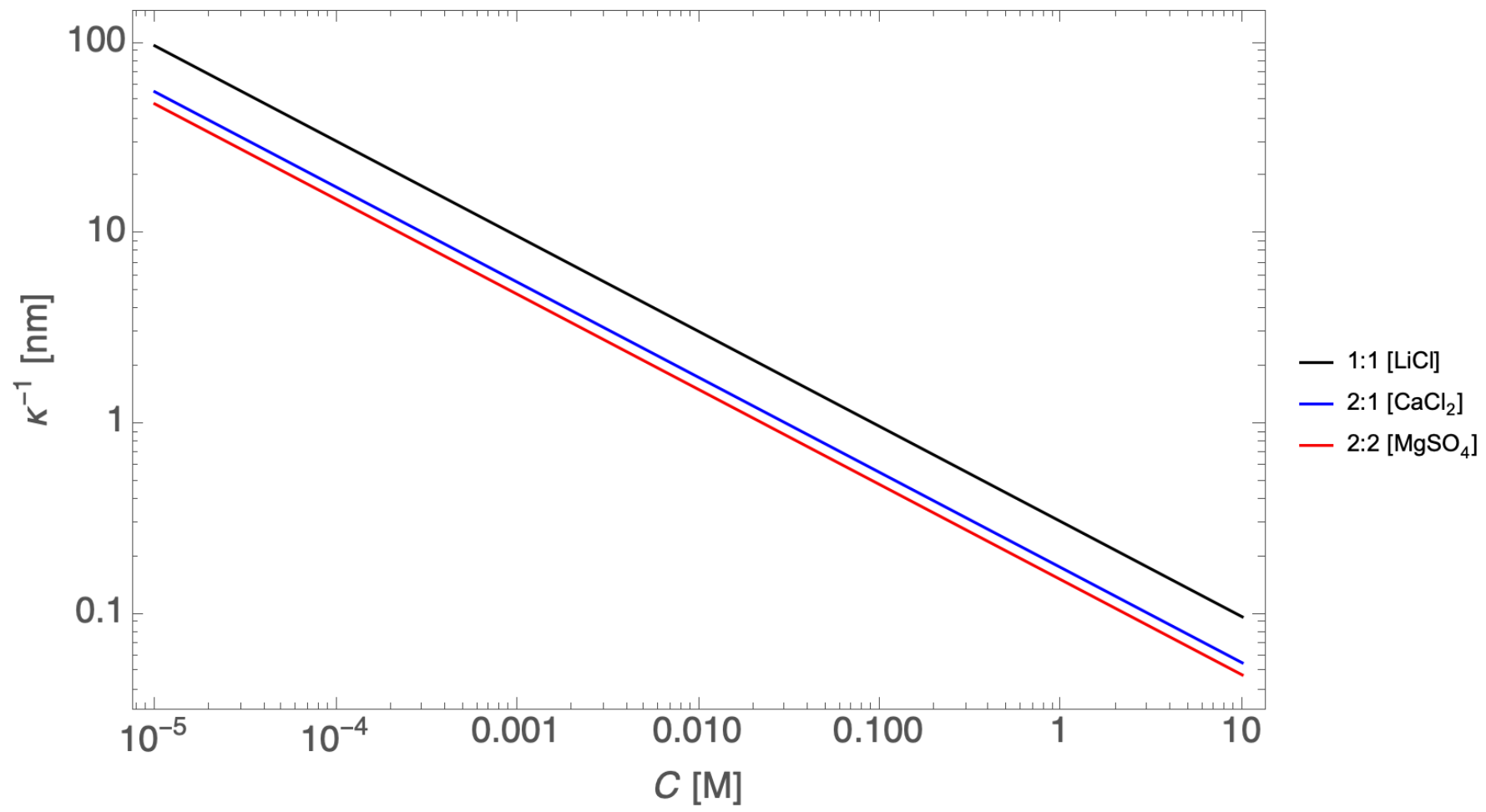

Fig S2 | Debye length for 1:1, 2:1, and 2:2 electrolytes.

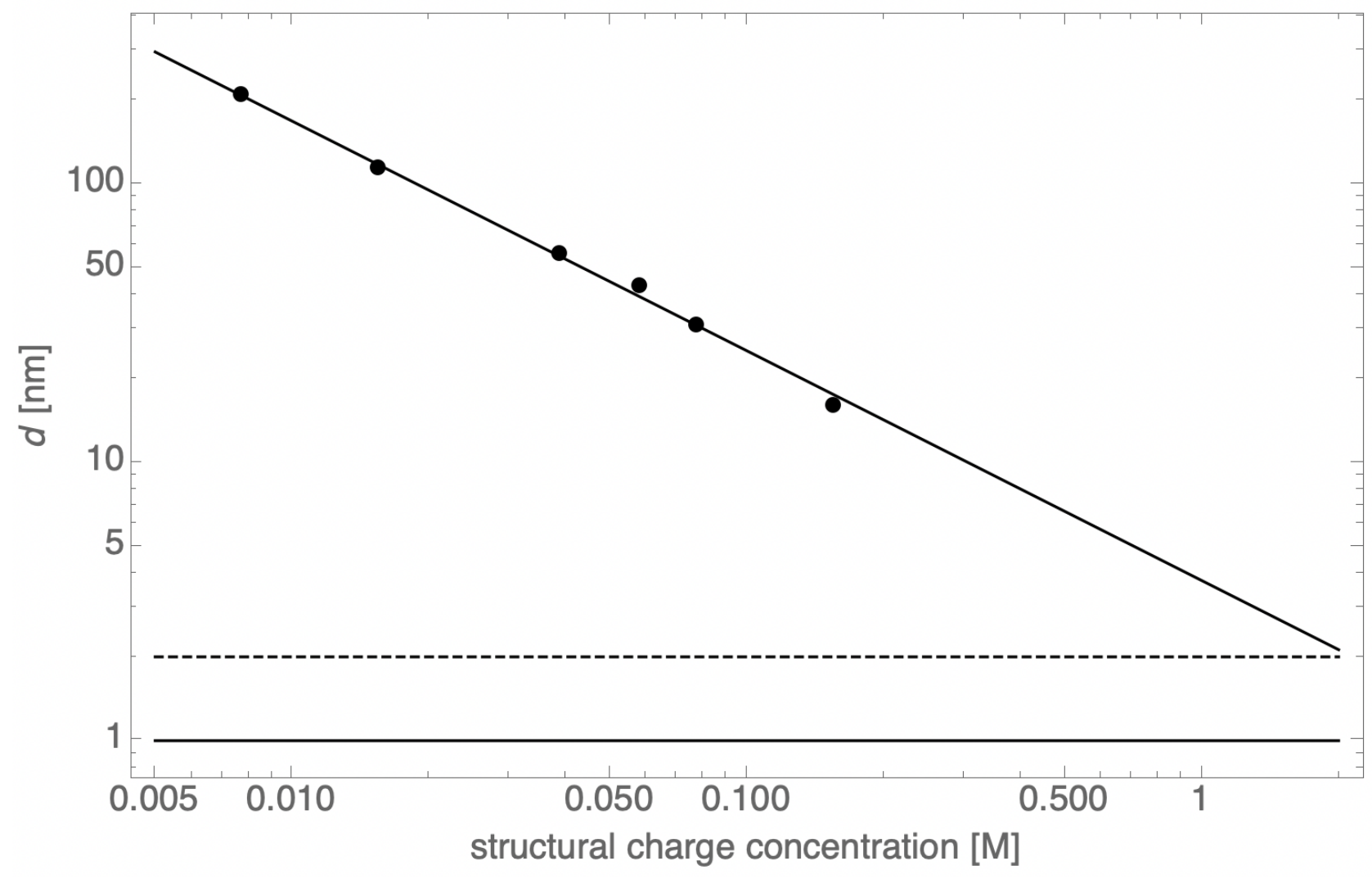

Fig S3 | Relationship between Mt charge concentration and *N* interlayer spacing in the absence of added electrolyte.

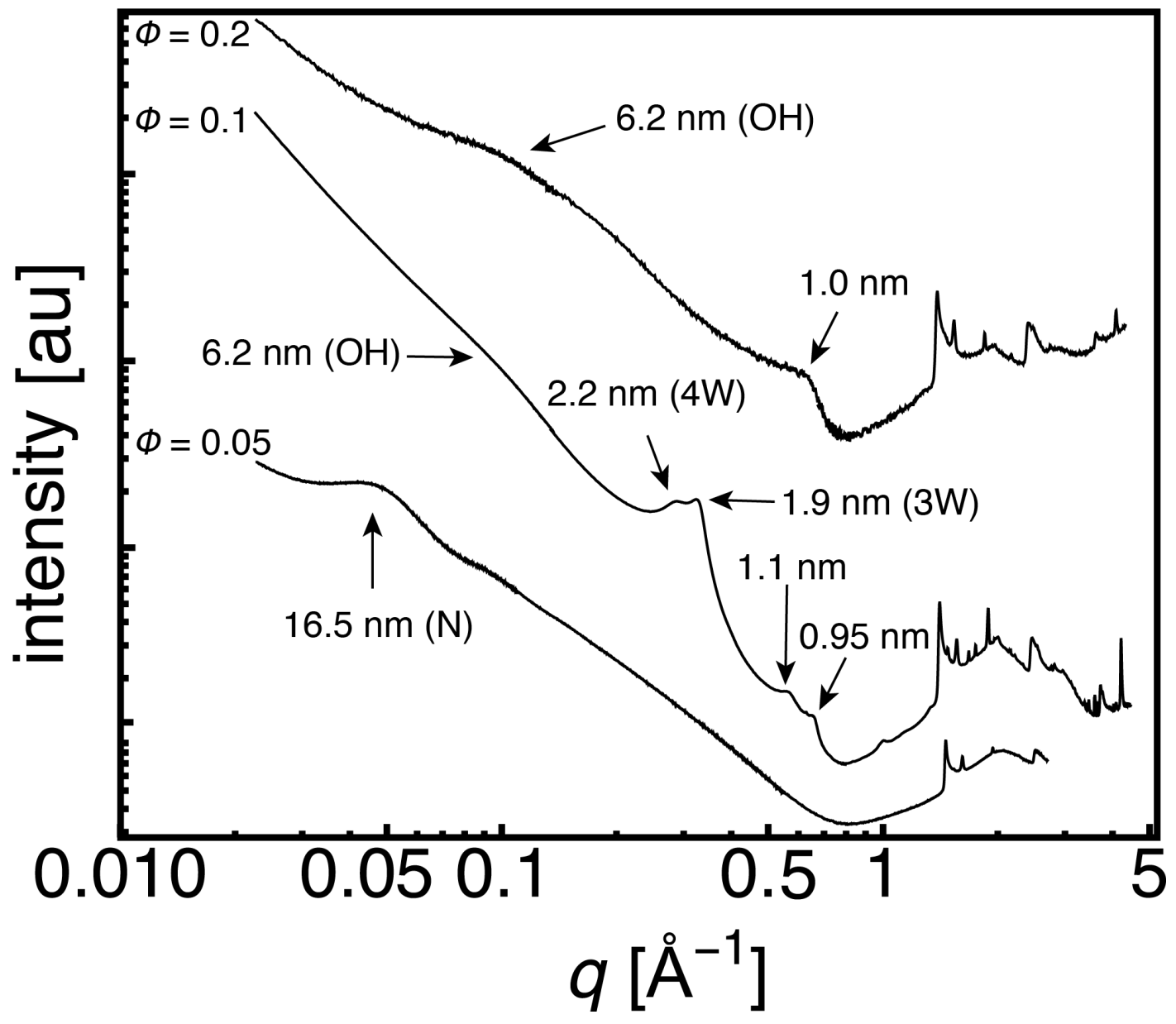

Fig S4 | 4W hydrate coexisting with 3W hydrate, which each exhibit different interfacial binding configurations as evidenced by different harmonic peaks

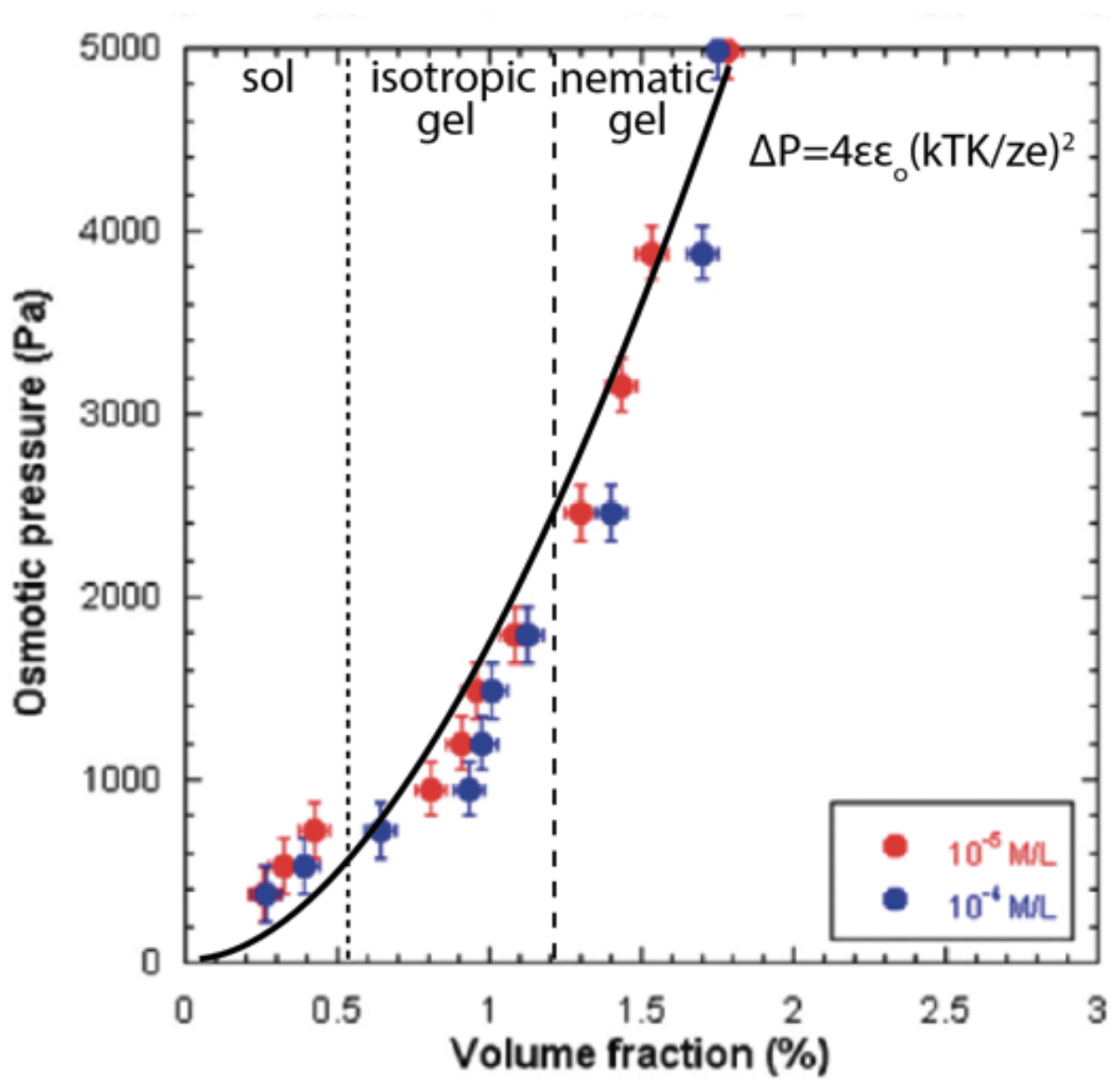

Fig S5 | Osmotic pressure of Na-Swy-2 as a function of mineral volume fraction, modified from[30] with fit to Eq. 11.

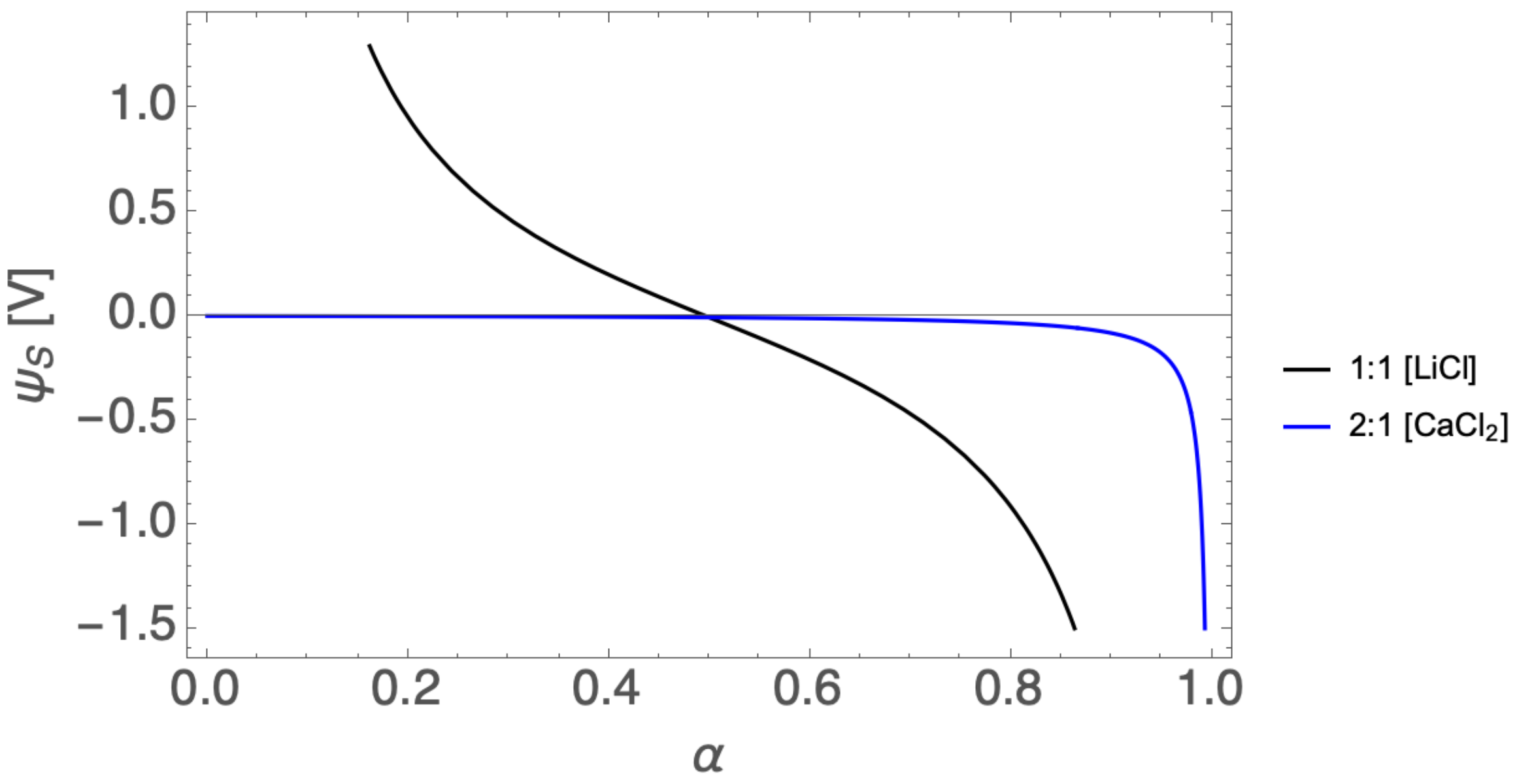

Fig S6 | Stern potential drop for LiCl and $CaCl_2$ electrolytes as a function of surface binding.

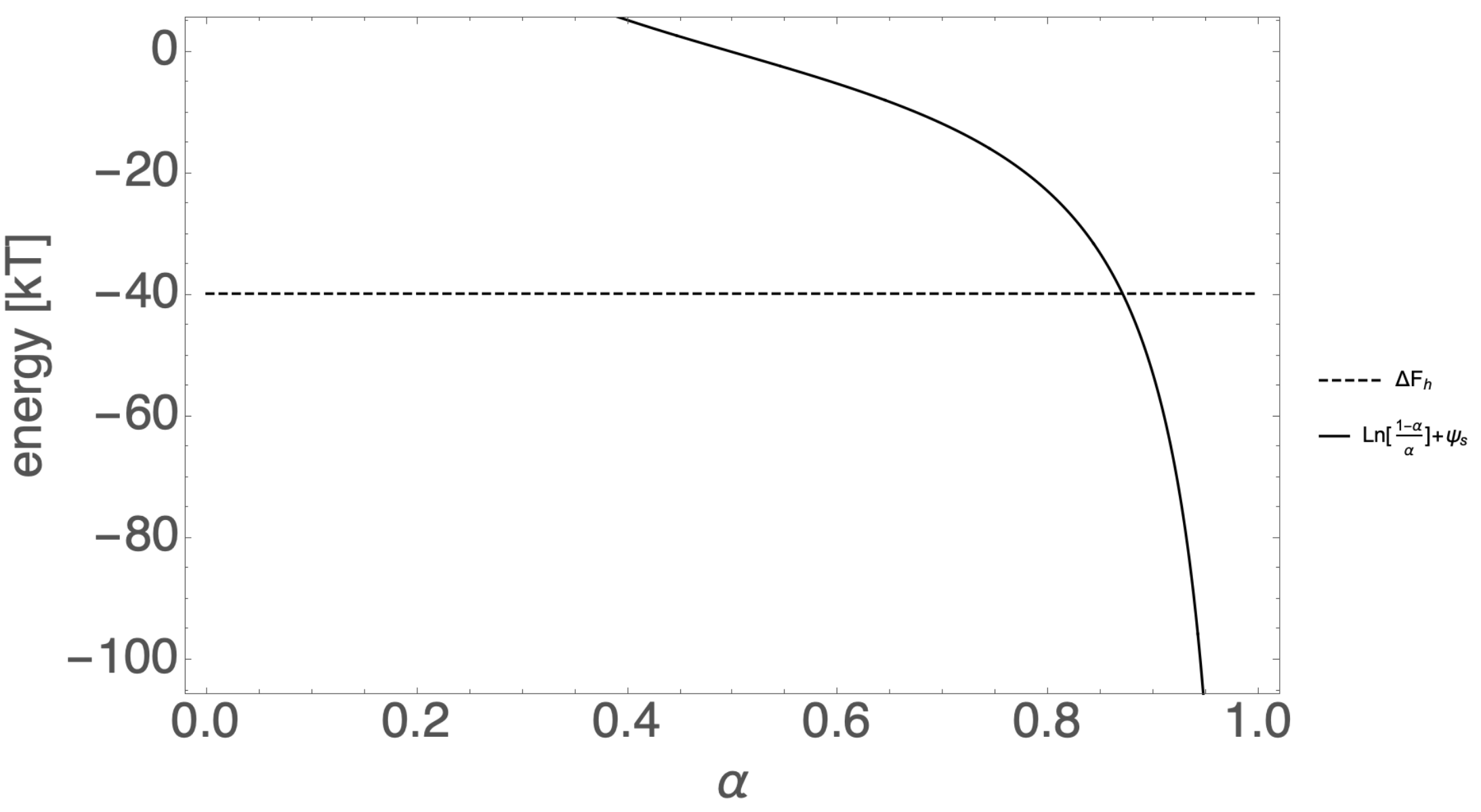

Fig S7 | Graphical solution of Eq. 14 for α.

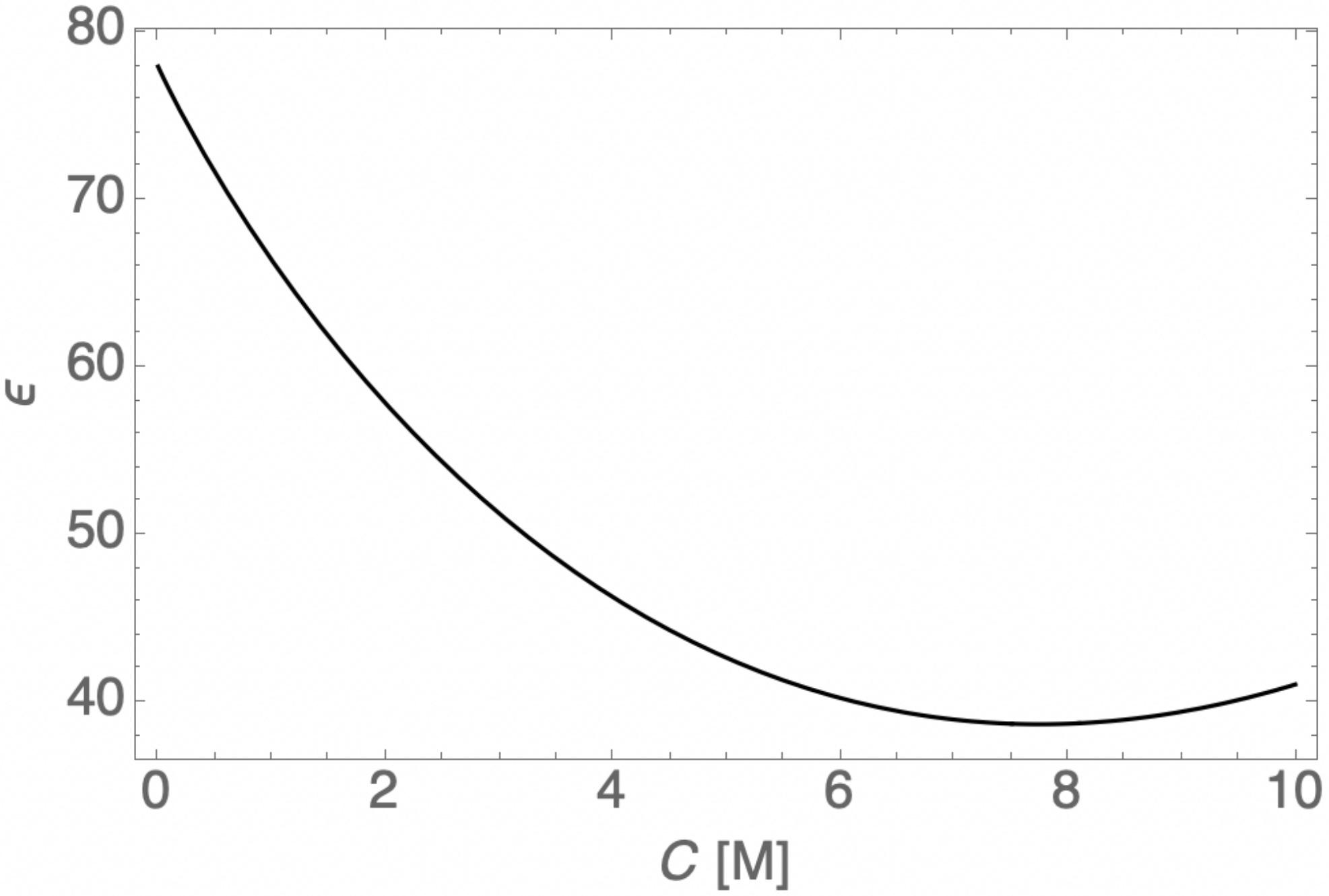

Fig S8 | Dielectric constant of bulk NaCl solution as a function of electrolyte concentration.

**Table S1. Hydration properties of divalent cations**[42]

| Ion | Ionic radius (Å) | $CN_w$ | $\Delta F_h$ ($kT/H_2O$) |
|---|---|---|---|
| $Mg^{+}$ | 0.86 | 6.0 | -120.2 |
| $Ca^{+}$ | 1.14 | 7.2 | -81.1 |
| $Sr^{+}$ | 1.32 | 8.0 | -67.2 |
| $Ba^{+}$ | 1.49 | 8.6 | -56.5 |

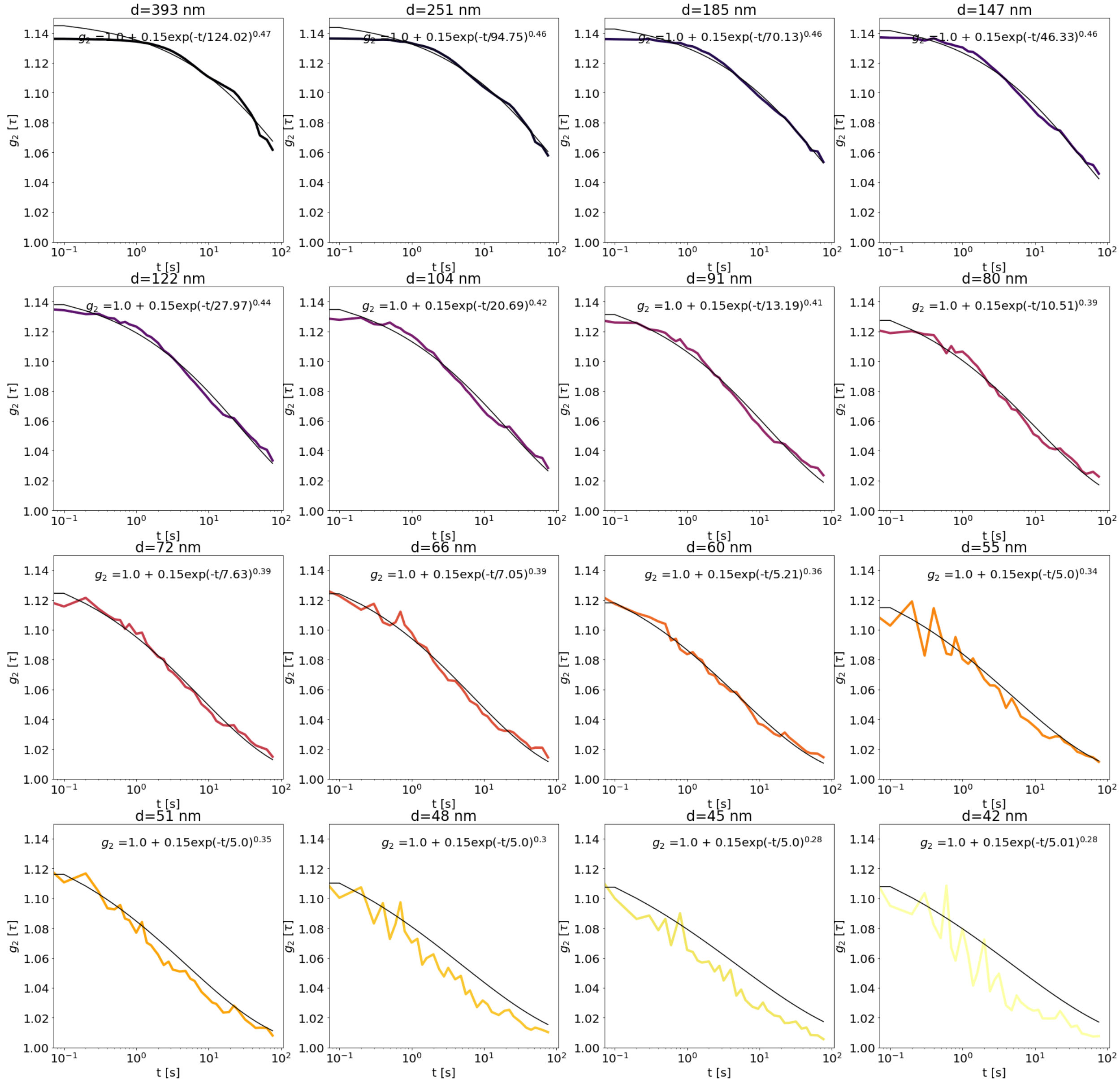

Fig S9 | Fits to one-time $g_2$ function, corresponding to data in Fig. 10c.

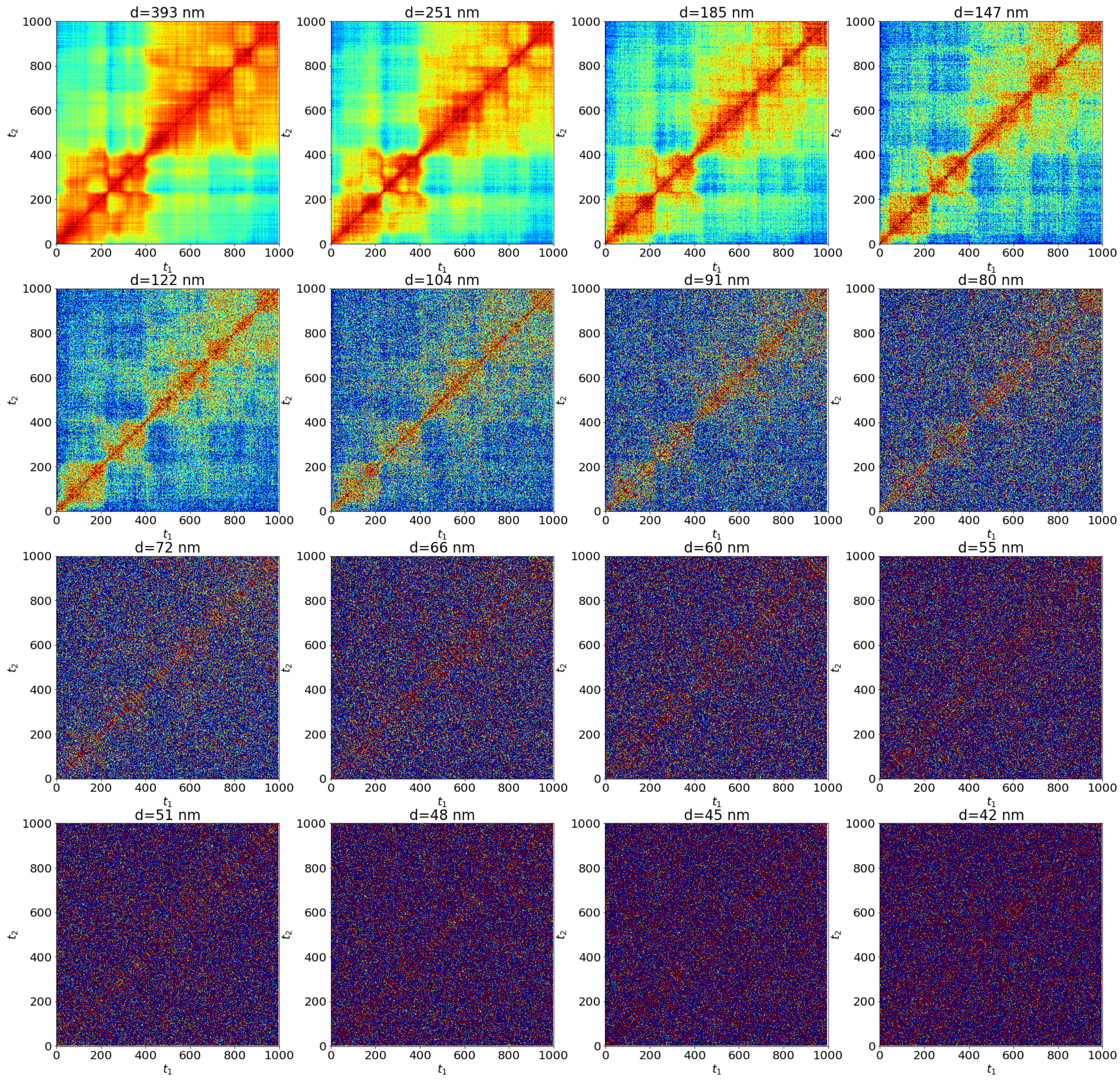

Fig S10 | Two-time correlation plots, corresponding to the data in Fig. 11d.